\documentclass[fleqn,usenatbib]{mnras}

\usepackage{times} 
\usepackage{listings}
\usepackage{graphics}
\usepackage{graphicx,xspace}
\usepackage{amsmath}
\usepackage{amssymb}
\usepackage{hyperref}
\usepackage{lscape}
\usepackage{rotating}
\usepackage{placeins}
\usepackage{natbib}
\usepackage{color}
\usepackage{txfonts}
\usepackage{mathptmx}
\usepackage{newtxtext,newtxmath}

\usepackage{subcaption}
\usepackage[dvipsnames]{xcolor}
\usepackage{caption}

\usepackage{grffile}
\usepackage{tabularx}
\usepackage{adjustbox}
\usepackage{array, makecell, cellspace}
\usepackage{multirow}
\usepackage{siunitx} 
\usepackage{tikz}

\newcommand{\ha}{H{\sc\,i}\xspace}

\title[The $M_{\rm HI}-M_{\star}$ relation  over the
last billion years]{MIGHTEE-\ha: The $M_{\rm HI}-M_{\star}$ relation over the last billion years}

\author[H. Pan et al.]{Hengxing Pan$^{1,2}$\thanks{E-mail:hpan@uwc.ac.za},
Matt J.~Jarvis$^{2,1}$,
Mario G. Santos$^{1,3}$,
Natasha Maddox$^{4,5}$,
Bradley S. Frank$^{3,6,7}$,
\newauthor
Anastasia A.~Ponomareva$^{2}$,
Isabella Prandoni$^{8}$, Sushma Kurapati$^{7}$, Maarten Baes$^{9}$,
\newauthor
Pavel E. Mancera Piña$^{10}$, Giulia Rodighiero$^{11,12}$, Martin J. Meyer$^{13}$, Romeel Davé$^{14}$, 
\newauthor
Gauri Sharma$^{1,15,16}$, Sambatriniaina H. A. Rajohnson$^{7}$, Nathan J. Adams$^{17}$, Rebecca A. A. Bowler$^{17}$, 
\newauthor
Francesco Sinigaglia$^{11,12}$, Thijs van der Hulst$^{18}$, Peter W. Hatfield$^{2}$, Srikrishna Sekhar$^{6,19}$ and
\newauthor
Jordan D. Collier$^{6,20,21}$
\\
$^{1}$Department of Physics and Astronomy, University of the Western Cape, Cape Town 7535, South Africa\\
$^{2}$Astrophysics, University of Oxford, Denys Wilkinson Buiding, Keble Road, Oxford OX1 3RH, UK\\
$^{3}$South African Radio Astronomy Observatory (SARAO), 2 Fir Street, Observatory, 7925, South Africa\\
$^{4}$School of Physics, H.H. Wills Physics Laboratory, Tyndall Avenue, University of Bristol, Bristol, BS8 1TL, UK\\
$^{5}$University Observatory, Faculty of Physics, Ludwig-Maximilians Universität, Scheinerstr. 1, 81679 Munich, Germany\\
$^{6}$The Inter-University Institute for Data Intensive Astronomy (IDIA), University of Cape Town, Private Bag X3, Rondebosch, 7701, South Africa\\
$^{7}$Department of Astronomy, University of Cape Town, Private Bag X3, Rondebosch 7701, South Africa\\
$^{8}$INAF - Istituto di Radioastronomia, Via P. Gobetti 101, 40129 Bologna, Italy\\
$^{9}$Sterrenkundig Observatorium, Universiteit Gent, Krijgslaan 281 S9, 9000 Gent, Belgium\\
$^{10}$Leiden Observatory, Leiden University, P.O. Box 9513, 2300 RA Leiden, The Netherlands\\
$^{11}$Dipartimento di Fisica e Astronomia, Università di Padova, Vicolo dell'Osservatorio, 3, I-35122, Padova, Italy\\
$^{12}$INAF--Osservatorio Astronomico di Padova, Vicolo dell'Osservatorio 5, I-35122, Padova, Italy\\
$^{13}$International Centre for Radio Astronomy Research (ICRAR), The University of Western Australia, 35 Stirling Hwy, Perth, WA 6009, Australia\\
$^{14}$Institute for Astronomy, Royal Observatory, Univ. of Edinburgh, Edinburgh EH9 3HJ, UK\\
$^{15}$Scuola Internazionale Superiore di Studi Avanzati, Trieste, Italy\\
$^{16}$INFN Sezione di Trieste, Italy\\
$^{17}$Jodrell Bank Centre for Astrophysics, Department of Physics and Astronomy, The University of Manchester, Manchester, M13 9PL, UK\\
$^{18}$Kapteyn Astronomical Institute, University of Groningen, Landleven 12, 9747AD Groningen, Netherlands\\
$^{19}$National Radio Astronomy Observatory, 1003 Lopezville Road, Socorro, NM 87801, USA\\
$^{20}$School of Science, Western Sydney University, Locked Bag 1797, Penrith, NSW 2751, Australia\\
$^{21}$CSIRO Astronomy and Space Science, PO Box 1130, Bentley, WA, 6102, Australia
}

\date{Accepted XXX. Received YYY; in original form ZZZ}

\pubyear{2022}

\begin{document}
\label{firstpage}
\pagerange{\pageref{firstpage}--\pageref{lastpage}}
\maketitle

\begin{abstract}
We study the $M_{\rm HI}-M_{\star}$ relation over the last billion years using the MIGHTEE-\ha sample. We first model the upper envelope of the $M_{\rm HI}-M_{\star}$ relation with a Bayesian technique applied to a total number of 249 \ha-selected galaxies, without binning the datasets, while taking account of the intrinsic scatter. We fit the envelope with both linear and non-linear models, and find that the non-linear model is preferred over the linear one with a measured transition stellar mass of $\log_{10}(M_\star$/$M_{\odot})$ = $9.15\pm0.87$, beyond which the slope flattens. This finding supports the view that the lack of \ha gas is ultimately responsible for the decreasing star formation rate observed in the massive main-sequence galaxies. For spirals alone, which are biased towards the massive galaxies in our sample, the slope beyond the transition mass is shallower than for the full sample, indicative of distinct gas processes ongoing for the spirals/high-mass galaxies from other types with lower stellar masses. We then create mock catalogues for the MIGHTEE-\ha detections and non-detections with two main galaxy populations of late- and early-type galaxies to measure the underlying $M_{\rm HI}-M_{\star}$ relation. We find that the turnover in this relation persists whether considering the two galaxy populations as a whole or separately. We note that an underlying linear relation could mimic this turnover in the observed scaling relation, but a model with a turnover is strongly preferred. Measurements on the logarithmic average of \ha masses against the stellar mass are provided as a benchmark for future studies.
\end{abstract}




\section{Introduction}
The relation between the mass of neutral atomic hydrogen (\ha) gas and stars in galaxies reveals the connection of star forming activity to their raw fuel, but this relation is not straightforward due to complex physical processes involved in the course of galaxy evolution. A comprehensive and accurate measurement of this relation is required to illuminate their interplay, and thus to help us better understand the physics of galaxy formation and evolution.

Over the past few decades, the direct detection of emission lines from the neutral hydrogen component of galaxies has been limited to the local Universe, or massive \ha systems, by the sensitivity of modern radio instruments, such as Parkes and Arecibo telescopes. Nonetheless, several studies have been conducted to investigate the \ha and stellar mass relation, benefiting from the \ha Parkes All-Sky (HIPASS) Survey  \citep{barnes2001h} and the Arecibo Legacy Fast ALFA (ALFALFA) survey \citep{giovanelli2005arecibo}.

In particular, exploring the upper envelope of the \ha and stellar mass ($M_{\rm HI}-M_{\star}$) relation has been one of the main means used to enlighten the processes of gas consumption and star formation. For example, both \cite{huang2012} and \cite{Maddox2015} have found an upper limit for \ha mass as a function of the stellar mass at high masses for \ha-selected samples. \cite{Maddox2015} and \cite{Parkash2018} suggest that this upper limit can be explained by a stability model in which the large halo spin of disk galaxies can stabilize the \ha disk and prevent it from collapsing and forming stars \citep{Obreschkow_2016}. In this scenario, the highest spin parameter is restrained by the amount of gas infall and tidal torque that haloes can experience during the proto-galactic growth, and therefore the gas fraction is linked to the specific angular momentum of galaxies in general \citep{Zoldan2018,Mancera2021}. As such this could also be related to the position of the galaxies with respect to the cosmic web, the filaments of which are presumably the source of the infalling gas \citep[see e.g.][]{Tudorache2022}. On the other hand, the underlying $M_{\rm HI}-M_{\star}$ relation has been investigated by \cite{Catinella_2010} and \cite{Parkash2018} for example, based on stellar mass-selected samples as \ha-selected samples tend to exclude \ha-poor galaxies resulting in measurements of high average (or median) \ha masses. They conclude that the observed flatness in the underlying $M_{\rm HI}-M_{\star}$ relation is due to the increasing fraction of gas poor early-type galaxies. All these studies indicate an increase in the \ha mass with stellar mass, but diverge at the high mass end owing to the effect of sample selection, limited statistics, or both.

Furthermore, there have been discoveries of flattening of the star formation rate (SFR)-$M_\star$ relation at high stellar masses from the local to high-z Universe \citep[e.g.][]{Noeske_2007,Erfanianfar_2015, Johnston2015, Leslie_2020, Fraser_McKelvie_2021}. The mechanisms responsible for this flattening remain under debate \citep[e.g.][]{Gavazzi_2015,Tacchella_2018,Popesso_2019,Cook_2020,Feldmann2020}, and can be broadly summarised as the lack of \ha gas versus the low conversion efficiency from \ha to stars, through the molecular hydrogen (H$_2$) phase. A thorough investigation into the link between \ha and the stellar mass can help to disentangle these two possible causes. Noticeably, the flattening of the SFR-$M_\star$ relation towards high masses resembles the upper limit found on the $M_{\rm HI}-M_{\star}$ relation.

At high redshifts, stacking approaches \citep[e.g.][]{Delhaize_2013, Rhee_2013, Healy_2019,Chowdhury_2020,Guo_2021,Sinigaglia2022, Bera_2022, Bera_2023} have been developed to break the barrier of the sensitivity limitation. However, in these stacking processes, one only measures the average properties of galaxies bearing the consequence of losing information about their intrinsic scatter, which is a key parameter to describe the shape of the distribution of \ha masses and hence the strength of the correlation between \ha and a second galaxy property, such as the stellar mass.

In addition, only the arithmetic operations (e.g. average) of the \ha fluxes are allowed for these stacking practices as the logarithmic operation cannot be done for negative fluxes that are influenced by the background noise, although an arithmetic average would be sufficient if we were just interested in measuring the cosmic \ha density. In terms of scaling relations, there are notable differences in the means and standard deviations between arithmetic and logarithmic operations of \ha masses mostly due to the different contribution of the low mass samples \citep{Rodr_guez_Puebla_2020,Saintonge2022}. For \ha-selected samples, the logarithmic average (or median) can adequately trace the main distribution, and is preferred in the literature \citep{Cortese2011, huang2012, Maddox2015, Parkash2018}. Therefore, it will add further complexities to a fair comparison of measured scaling relations between stacked samples and direct detections, based on different statistics. Above all, these approaches require binning the datasets in a second galaxy property, and it could be troublesome to determine the binning width when the sample size is small.

With the MeerKAT radio telescope and the future SKA, we are now entering a new era of radio astronomy. The MeerKAT International GHz Tiered Extragalactic Exploration \citep[MIGHTEE; ][]{Jarvis2016} is one of the large survey projects that is underway with MeerKAT, and will cover 20 square degrees over the four best-studied extragalactic fields observable from the southern hemisphere to $\mu$Jy sensitivity at GHz frequencies. The MIGHTEE-\ha Early Science project has already allowed us to reach $M_{\rm HI}\lesssim10^7 M_{\odot}$ in the local Universe, and $M_{\rm HI}\sim10^9 M_{\odot}$ up to $z=0.084$, with higher H{\sc i}-mass galaxies observable out to the lower-frequency end of the L-band window at $z\sim 0.6$ \citep{maddox2021}. Furthermore, the combination of this depth and high spatial resolution, resulting in essentially no source confusion compared to single-dish surveys, and the depth of the ancillary data, allowing us to reach stellar masses of $\sim 10^6$\,M$_{\star}$, makes this a unique data set to investigate the $M_{\rm HI}-M_{\star}$ scaling relation over a large baseline in \ha and stellar mass.

In this paper, we first use a Bayesian technique for modelling the upper envelope of the $M_{\rm HI}-M_{\star}$ scaling relation consistently without binning the datasets from the MIGHTEE-\ha catalogue, while taking account of the intrinsic scatter, based on our previous work \citep{Pan2020, Pan_2021}. This technique employs fluxes of \ha emission line as measurables that can naturally account for the thermal noise from the radio receiver on the linear scale, while the intrinsic scatter of galaxy properties may be better described on the logarithmic scale. This is non-trivial as the propagation of uncertainties measured from the linear to logarithmic scale must rely on an approximation which breaks down when the signal to noise ratio is low. If we instead measure the scaling relation in flux space, this issue does not exist. Our approach can also mitigate low number statistics without the binning strategy. We note that this technique can be easily adjusted and applied to measure other \ha scaling relations and the \ha mass function directly above or below the detection threshold. We then create mock MIGHTEE-\ha galaxies to quantify the \ha-selection effect and measure the underlying $M_{\rm HI}-M_{\star}$ relation by dividing the mock samples into late- and early-type galaxies.

This paper is organised as follows. We describe our MIGHTEE-\ha and the ancillary data in Section \ref{sec:data}, and the Bayesian technique in Section \ref{sec:method}. We present our main results in Section \ref{sec:results}, and conclude in Section \ref{sec:conclusions}. We use the standard $\Lambda$CDM cosmology with a Hubble constant $H_{0}=67.4$ km$\cdot$s$^{-1}\cdot {\rm Mpc}^{-1}$, total matter density $\Omega_{\rm m}=0.315$ and dark energy density $\Omega_{\Lambda}=0.685$ \citep{planck2020}, and AB magnitudes throughout.

\section{Data}
\label{sec:data}

\subsection{MIGHTEE-\ha}
\label{sec:HIcubes}

MIGHTEE-\ha is the \ha emission project within the MIGHTEE survey, and is described in detail by \cite{maddox2021}. The MIGHTEE–\ha Early Science data were collected between mid-2018 and mid-2019, in L–band (900 $< \nu <$ 1670 MHz), with a spectral resolution of 208 kHz over two well-studied fields: COSMOS and XMMLSS. The visibilities were
processed with the IDIA Pipeline\footnote{\url{https://idia-pipelines.github.io}}: processMeerKAT. This pipeline does full-polarisation calibration on MeerKAT data including automated flagging. The spectral line imaging was carried out using the CASA task TCLEAN (robust=0.5), and the continuum subtraction was undertaken in both the visibilities and imaging planes using the standard CASA routines UVSUB and UVCONTSUB. The effect of direction-dependent artefacts was reduced by a per-pixel median filtering operation. The full data reduction pipeline for MIGHTEE-\ha is described by Frank et al. (in prep). Key parameters of the processed Early Science data are listed in Table \ref{tab1}. We note that this combination of depth and resolution is unique for a blind \ha survey.

\begin{table}
        \small
        \centering
        \caption{Key parameters of the MIGHTEE-\ha Early Science data.}
        \label{tab1}
        \begin{tabular}{l l}
                \hline
                \hline
                Survey area & ${\sim}$1.5 deg$^{2}$ (COSMOS) \\
                & ${\sim}$3 $\times$ 1.2 deg$^{2}$ (XMMLSS) \\
                Integration time  & ${\sim}16$ hrs (COSMOS) \\
                & 3 $\times$ 12 hrs (XMMLSS)\\
                Velocity resolution &   44.11 kms$^{-1}\,$ at $z = 0$ \\
                Synthesized Beam & $\ang{;;14.5} \times \ang{;;11}$ (COSMOS)\\
                & $\ang{;;12} \times \ang{;;10}$ (XMMLSS)\\
                $3\sigma$ \ha \ column density sensitivity & $4.05 \times 10^{19}$ atoms\,cm$^{-2}$\ (COSMOS) \\
                & $9.83 \times 10^{19}$ atoms\,cm$^{-2}$\ (XMMLSS)\\
                \hline
        \end{tabular}
\end{table}

\subsection{\ha flux}
\label{sec:hiflux}

We employ the Cube Analysis and Rendering Tool for Astronomy \citep[CARTA;][]{Comrie2021} for visual source finding, then we extract a cubelet centred on all detected sources. We smooth the cubelet, and clip it at a 3$\sigma$ threshold as a mask for removing the noise, then apply the mask to the cubelet with original resolution. We then clip out by hand any remaining noise peaks and integrate the flux densities over the frequency channels to make moment-0 maps. The total flux is calculated by summing all flux densities over the spatial pixels. We obtain uncertainties on the bright and faint sources varying from  5\% to 20\% of their \ha fluxes roughly \citep[see][for full details]{Ponomareva_2021}.

The \ha mass under the optically thin gas assumption can be calculated via
\begin{equation}
    M_{\rm HI} = 2.356 \times 10^5 D^2_L(1+z)^{-1} S,
	\label{eq:factor}
\end{equation}
where $M_{\rm HI}$ is the \ha mass in solar masses, $D_L$ is the luminosity distance in Mpc, and $S$ is the integrated flux in Jy\,km\,s$^{-1}$ \citep{meyer2017tracing}. We note that our technique works on the \ha flux space directly rather than the mass space, and this equation is only needed for our technique to predict the flux $S$ when the \ha mass is modelled by the $M_{\rm HI}-M_{\star}$ relation in Section~\ref{sec:model}.

\subsection{Ancillary data}
\label{sec:stmass}

All MIGHTEE fields are covered by various multi-wavelength photometric and spectroscopic surveys ranging from X-ray to far-infrared bands \citep[e.g.][]{Cuillandre2012, McCracken2012, Jarvis_2013, Aihara_2017, Aihara2019}. We measure the magnitudes of the sample galaxies by extracting the flux within an elliptical aperture defined in the $g$–band, and we apply this aperture to the $urizYJHK_s$–bands. Based on independent, repeat measurements of several galaxies, the photometry is accurate to $\sim$0.015 mags. We then employ the Spectral Energy Distribution (SED) fitting code {\sc LePhare} \citep{Ilbert2006} for deriving the stellar properties of the galaxies, such as stellar mass, stellar age and star formation rate, and the uncertainty on the stellar mass is conservatively assumed to be $\sim$0.1 dex, due to assumptions made on galaxy metallicity, star formation history, initial mass function (IMF), etc. in the SED fitting process \citep{Adams_2021, maddox2021}. In particular, the star formation histories use \cite{BC03} stellar synthesis models including templates with either a constant star formation history or an exponential star formation history. For the exponential star formation history, there are a few different characteristic timescales for the exponent ($\tau$) ranging from $\tau$ = [0.1, 0.3, 1, 2, 3, 5, 10, 15, 30] Gyrs. For each template, there are also 57 different ages from 0.01 Gyr up to the age of the universe.

Supplemented with such a rich ancillary data set, we are in an excellent position to study \ha galaxies from various perspectives, and understand the links between \ha gas and other key galaxy properties such as colour, SFR, and stellar mass, in order to gain a complete picture of the diverse galaxy populations as they evolve across the cosmic time.

\begin{figure*}
  \centering
  \begin{subfigure}[s]{0.5\textwidth}
    \includegraphics[width=\columnwidth]{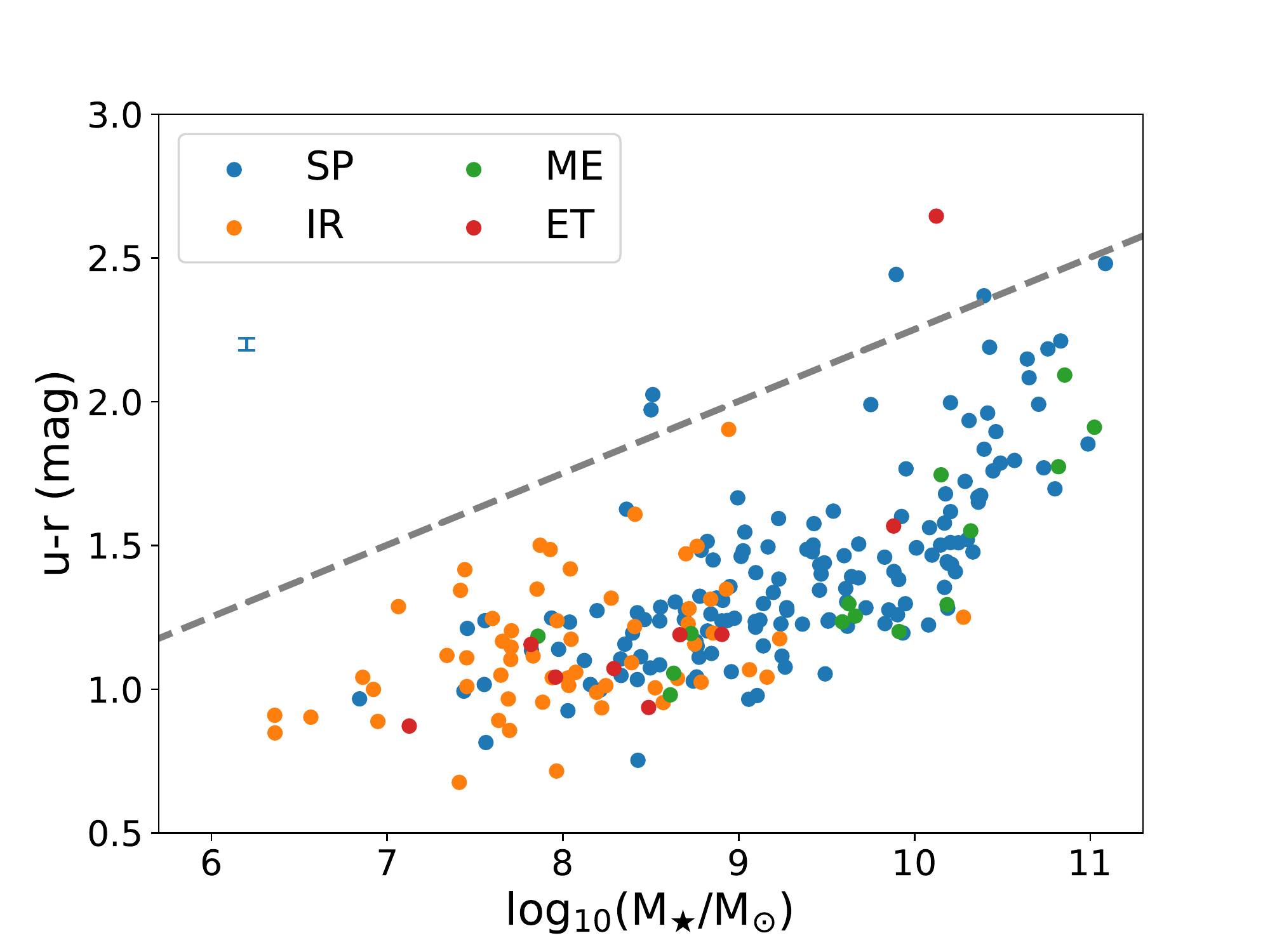}
  \end{subfigure}%
  \hfill
  \begin{subfigure}[s]{0.5\textwidth}
    \includegraphics[width=\columnwidth]{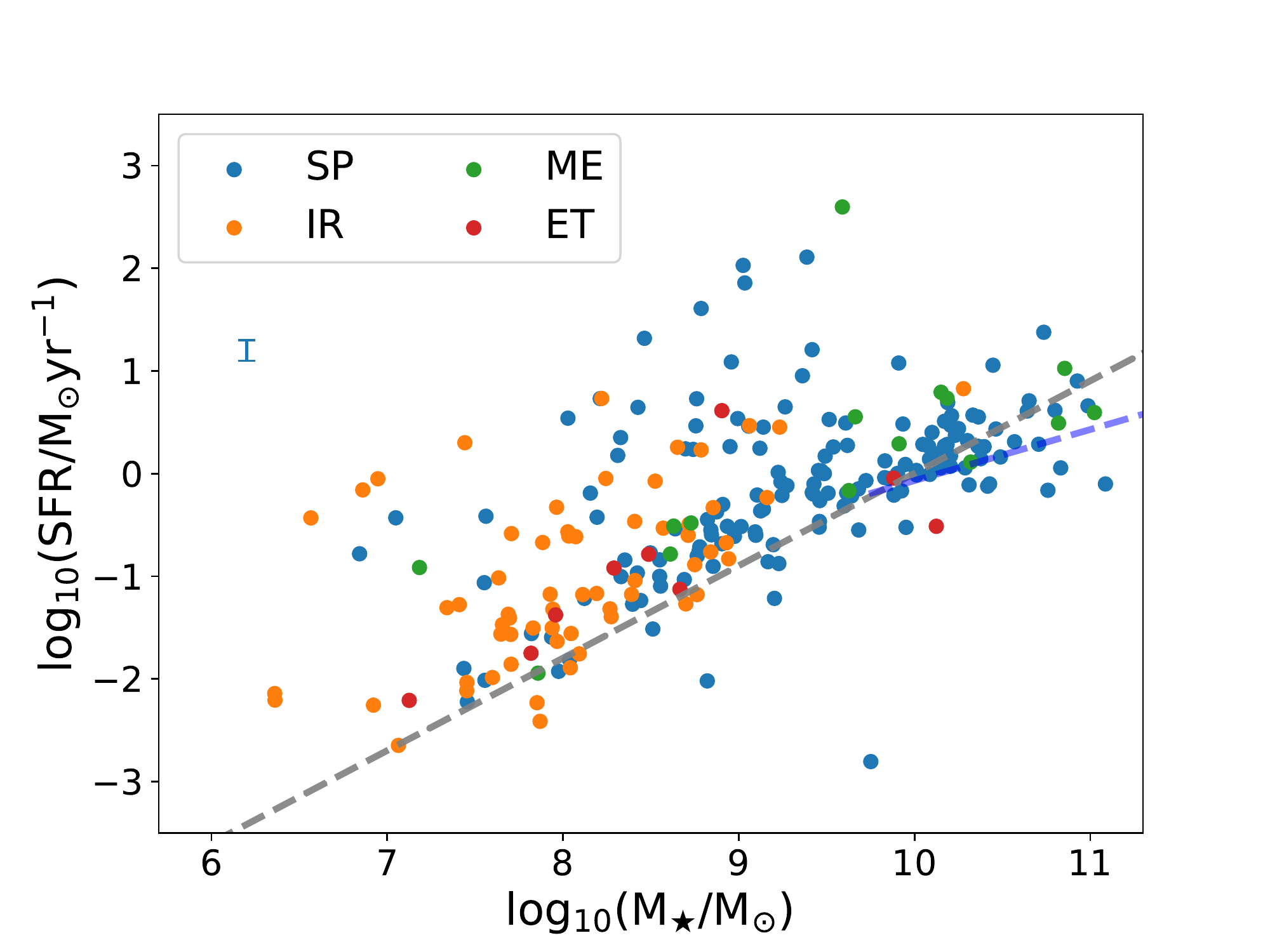}
  \end{subfigure}%
  \caption{Colour (left) and SFR (right) against the stellar mass. Measurements are color-coded by the morphologies of galaxies classified as spirals (SP), irregulars (IR), mergers (ME), and ellipticals (ET). The grey dashed line in the left panel is the upper boundary of the green valley galaxies from \protect\cite{Schawinski2014}, while in the right panel the grey and blue dashed lines show the ridge lines of the main sequence of star forming galaxies from \protect\cite{Peng2010} and \protect\cite{Speagle_2014}, who used the same IMF \citep{Chabrier} and stellar population synthesis models \citep{BC03}, as used in this paper. 1$\sigma$ uncertainties on the galaxy $u-r$ colour and SFR are illustrated with blue error bars in the top left corners respectively.}
  \label{fig:color-SFR-mstar}
\end{figure*}

\subsection{Morphological classification}
\label{sec:morpho}

We classify our \ha detections into four samples based on their optical morphology. In total, we have a sample of 276 \ha detections. By removing the objects outside the deep imaging footprint of the Hyper SuprimeCam Subaru Strategic Program \citep[HSC-SSP][]{Aihara_2017} and the ones without a classified type, we have a total number of 249 galaxies including 161 spiral galaxies (SP), 64 irregulars (IR), 15 mergers (ME), and 9 elliptical galaxies (ET). Details of the galaxy morphology classification are described in \cite{Rajohnson_2022}. We note that many of the galaxies classified as irregular are in fact very low mass, and thus could alternatively be classified as dwarf galaxies. We do not differentiate early and late-stage mergers due to their small sample size, and label all of them as ME in our analysis.

In Figure~\ref{fig:color-SFR-mstar}, we show the colour- and SFR-stellar mass diagrams colour-coded by their morphologies in the left and right panels, respectively. We draw the upper boundary of the green valley galaxies from \cite{Schawinski2014} as the grey dashed line on the colour-stellar mass diagram, and ridge lines of the main sequence of star forming galaxies from \protect\cite{Peng2010} and \protect\cite{Speagle_2014} as grey and blue dashed lines, respectively. These demonstrate that our \ha-selected sources are mostly settled in the blue cloud and green valley, and largely distributed above the main sequence of star forming galaxies, with very few being red (or passive) galaxies. The irregular and spiral galaxies dominate at the low and high mass ends, respectively, with considerable overlapping area at the intermediate mass range. This feature motivates us to separate the \ha-selected galaxies based on their morphology, and investigate the dominant sample of spirals, in addition to the H{\sc i}-selected sample as a whole.

It is worth noting that the SFR-$M_\star$ relation predominantly follows a power law at low stellar masses, and flattens at $M_\star \gtrsim 10^{10}M_\odot$ as expected for the main sequence galaxies \citep[e.g.][]{Lee_2015, Schreiber2015,Saintonge_2016,Leslie_2020}, even with our limited sample size from the Early Science data.

\section{Bayesian analysis}
\label{sec:method} 

\subsection{Bayesian framework}

Our technique is established on Bayes’ theorem
\begin{equation}
\centering
\mathcal{P}(\Theta|D,H) =\frac{\mathcal{L}(D|\Theta,H) \Pi(\Theta|H)}{\mathcal{Z}(D|H)},
\label{eqn:bayes}
\end{equation}
where $\mathcal{P}$ is the posterior distribution of the model parameters $\Theta$, given
the available data $D$ and a model $H$. \textbf{$\mathcal{L}$} is the likelihood  of the data $D$ given parameter values and the model, and $\Pi$ is
the prior knowledge of our prejudices about the values of the model parameters. $\mathcal{Z}$ is the Bayesian evidence, which can be thought of as a normalization factor and can be expressed as an integral of $\mathcal{L}$ and $\Pi$ over a $n$-dimensional parameter space $\Theta$,
\begin{equation} 
\mathcal{Z}(D|H)  = \int \mathcal{L}(D|\Theta,H) \Pi(\Theta|H) \textrm{d}^n\Theta,
\label{eqn:Z} 
\end{equation} 
and in addition it crucially facilitates model selection between different models when their evidences are compared quantitatively, as the evidence is the probability of the data given a model after all the free parameters are marginalized over. The
difference in the log-evidence, $\Delta\ln(\mathcal{Z_B}) = \ln(\mathcal{Z_B})$ - $\ln(\mathcal{Z_A})$, known as
the Bayes factor, is commonly used to interpret how much better Model B is compared to A, providing a fair way of discriminating between models with differrent numbers of parameters by penalising models that are too flexible. We follow the criteria in \cite{Malefahlo_2021}, where $\Delta\ln(\mathcal{Z}) < 1$ is "not significant", $1 < \Delta\ln(\mathcal{Z}) < 2.5$ is "significant", $2.5 < \Delta\ln(\mathcal{Z})< 5$ is "strong", and $\Delta\ln(\mathcal{Z}) > 5$ is "decisive".

We use {\sc Multinest} \citep{Feroz_2009,Buchner_2014}, an efficient and robust Bayesian inference tool for cosmology and particle physics, to sample the parameter
space and explore the full posterior distribution for parameter estimation and the evidence for Bayesian model comparison.

\subsection{$M_{\rm HI} - M_{\star}$ models}
\label{sec:model}

We fit two $M_{\rm HI} - M_{\star}$ models: linear and non-linear models to the data in their logarithmic space, given that both have been used previously \citep[e.g.][]{Maddox2015,Parkash2018}. First, we model the logarithmic average of $M_{\rm HI}$ as a linear function of $\log_{10}(M_{\star})$ as follows
\begin{equation}
    \langle\log_{10}(M_{\rm HI})\rangle = \alpha [\log_{10}(M_{\star})-10]+\beta,
    \label{eq:model-A}
\end{equation}
where $\alpha$ and $\beta$ are the free parameters corresponding to the slope, and intercept at $M_{\star}=10^{10} M_\odot$. We note this single power law relation as "\textbf{Model A}".

For the non-linear relation, we use the double power law relation:
\begin{equation}
    \langle\log_{10}(M_{\rm HI})\rangle =\log_{10}\left( \frac{M_0}{\left(\frac{M_\star}{M_{\rm tr}}\right)^a + \left(\frac{M_\star}{M_{\rm tr}}\right)^b}\right),
    \label{eq:model-B}
\end{equation}
where $M_{\rm tr}, M_0,  a , b$ are the free parameters to be fitted for. $M_{\rm tr}$ indicates the transition stellar mass, and $M_0$ is a value along the ordinate at $M_\star = M_{\rm tr}$ where we have $\langle\log_{10}(M_{\rm HI})\rangle = \log_{10}(M_0/2)$; $a$ and $b$ determine the low- and high-mass slopes of the scaling relation. We denote this double power law relation as "\textbf{Model B}", i.e. our non-linear model. When $a=b$, Eq.~\eqref{eq:model-B} is equivalent to Eq.~\eqref{eq:model-A}.

\subsection{Likelihood}
The relationship between \ha and stellar mass of galaxies cannot be fully described by a single variable function, no matter which model we use. We actually require a bivariate distribution function to capture the whole picture of the $M_{\rm HI} - M_{\star}$ relation. Without loss of generality, if we assume the Model A or B supplemented with an intrinsic scatter $\sigma_{\rm HI}$ is good enough to describe this relation for our relatively small sample, then the probability of having a \ha mass $(M_{\rm HI})$ at a given stellar mass $(M_{\star})$ follows,
\begin{equation}
    P(M_{\rm HI}|M_\star) = \frac{1}{\sqrt{2\pi}\sigma_{\rm HI}} e^{-\frac{1}{2}\left(\frac{\log_{10}(M_{\rm HI})-\langle\log_{10}(M_{\rm HI})\rangle}{\sigma_{\rm HI}}\right)^2}.
    \label{eq:chimf3}
\end{equation}
We take the intrinsic scatter $\sigma_{\rm HI}$ as an additional free parameter for our Models A and B. This Gaussian form of distribution function can be replaced with a Schechter function or any other forms if required.

With this conditional \ha mass distribution, the probability of having a measured flux, $S_{\rm m}$, for a single source can be expressed as
\begin{equation}
P(S_{\rm m}|M_\star) = \int d M_{\rm HI}P(M_{\rm HI}|M_\star) P_n(S_{\rm m}-S(M_{\rm HI})),
\label{eq:probability}
\end{equation}
where $P_n$ follows the noise distribution of $\rm Normal(0,\; \sigma_{\rm n})$, and $S(M_{\rm HI})$ is given by the Eq.~\eqref{eq:factor}.

The likelihood of all the sources having the measured fluxes, given the model and known stellar masses, is given by
\begin{equation}
    \begin{aligned}
    \mathcal{L} \propto  \prod_{\rm source} P(S_{\rm m} &|M_\star, {\rm Model\;A} (\alpha,\beta, \sigma_{\rm HI}) ,\\
    \rm or\; &|M_\star, {\rm Model\;B}(a, b, \sigma_{\rm HI}, M_0, M_{\rm tr})) .
	\label{eq:likeli}
    \end{aligned}
\end{equation}

By maximizing Eq.~\eqref{eq:likeli}, we obtain the best fitting $M_{\rm HI} - M_{\star}$ relation with an estimate of the intrinsic scatter for the given sample.

\subsection{Priors}
\label{sec:priors}

Priors are our background knowledge of the model parameters, and thus define the sampled parameter space. A uniform prior distribution provides an equal weighting of the input parameter space, and is preferred in general if this prior distribution is not known well. We assign uniform prior probability distributions to $\alpha$, $\beta$ and $\sigma_{\rm HI}$ for our Model A. For Model B, we assign uniform distributions to $a, b$ and  $\sigma_{\rm HI}$, and adopt uniform logarithmic priors for $M_0$ and $M_{\rm tr}$. All of these priors are listed in Table~\ref{tab:priors}.

\begin{table}
	\centering
	\caption{Priors of the Models A and B for the \ha and stellar mass relation.}
	\label{tab:priors}
	\begin{tabular}{cll}
        \hline
	\hline
	Model & Parameter  & Prior Probability Distribution \\
	\hline
          & $\alpha $            & uniform $\in [-2.5, 2.5] $    \\
	  A   & $\beta$           & uniform $\in [7, 12] $    \\
	      & $\sigma_{\rm HI}$    & uniform $\in [0, 2] $    \\
    \hline
         &  $a$          & uniform $\in [-2, 0] $    \\
         &  $b$           & uniform $\in [-0.5, 1.5] $    \\
	  B  & $\sigma_{\rm HI}$            & uniform $\in [0, 2] $    \\
	     & $\log_{10}(M_0)$         & uniform $\in [7, 12] $       \\
	     & $\log_{10}(M_{\rm tr})$      & uniform $\in [7, 12] $       \\
    \hline
	\label{tab:priors}
	\end{tabular}
\end{table}

\renewcommand{\arraystretch}{1.2}
\begin{table}
	\centering
	\caption{The best fitting parameters of the observed $M_{\rm HI}-M_{\star}$ relation with Models A and B for our MIGHTEE-\ha-selected samples at $0<z<0.84$. The values listed are those with the maximum likelihood from our fitting.}
	\label{tab:post-hi-AB}
 	\begin{adjustbox}{max width=\columnwidth}
	\begin{tabular}{clrr}
        \hline
	\hline
	Model & Parameter  & Spirals & Full Sample \\
	\hline
          & $\alpha $               & 0.278$\pm$0.040 &  0.387$\pm$0.027   \\
	  A   & $\beta$               & 9.649$\pm$0.043 & 9.693$\pm$0.039  \\
	      & $\sigma_{\rm HI}$     & 0.44$\pm$0.02    & 0.46$\pm$0.02   \\
    \hline
         &  $a$                     & -0.523$\pm$0.164 & -0.672$\pm$0.133   \\
         &  $b$                     & 0.022$\pm$0.429  & -0.035$\pm$0.194    \\
	  B  & $\sigma_{\rm HI}$      & 0.440$\pm$0.024   & 0.435$\pm$0.020   \\
	     & $\log_{10}(M_0)$        & 9.869$\pm$0.365  & 9.771$\pm$0.327    \\
	     & $\log_{10}(M_{\rm tr})$ & 9.52$\pm$1.56     & 9.15$\pm$0.87      \\
    \hline
	\end{tabular}
	\end{adjustbox}
\end{table}

\section{Results}
\label{sec:results}

\begin{figure*}
  \centering
  \begin{subfigure}[s]{0.5\textwidth}
    \includegraphics[width=\columnwidth]{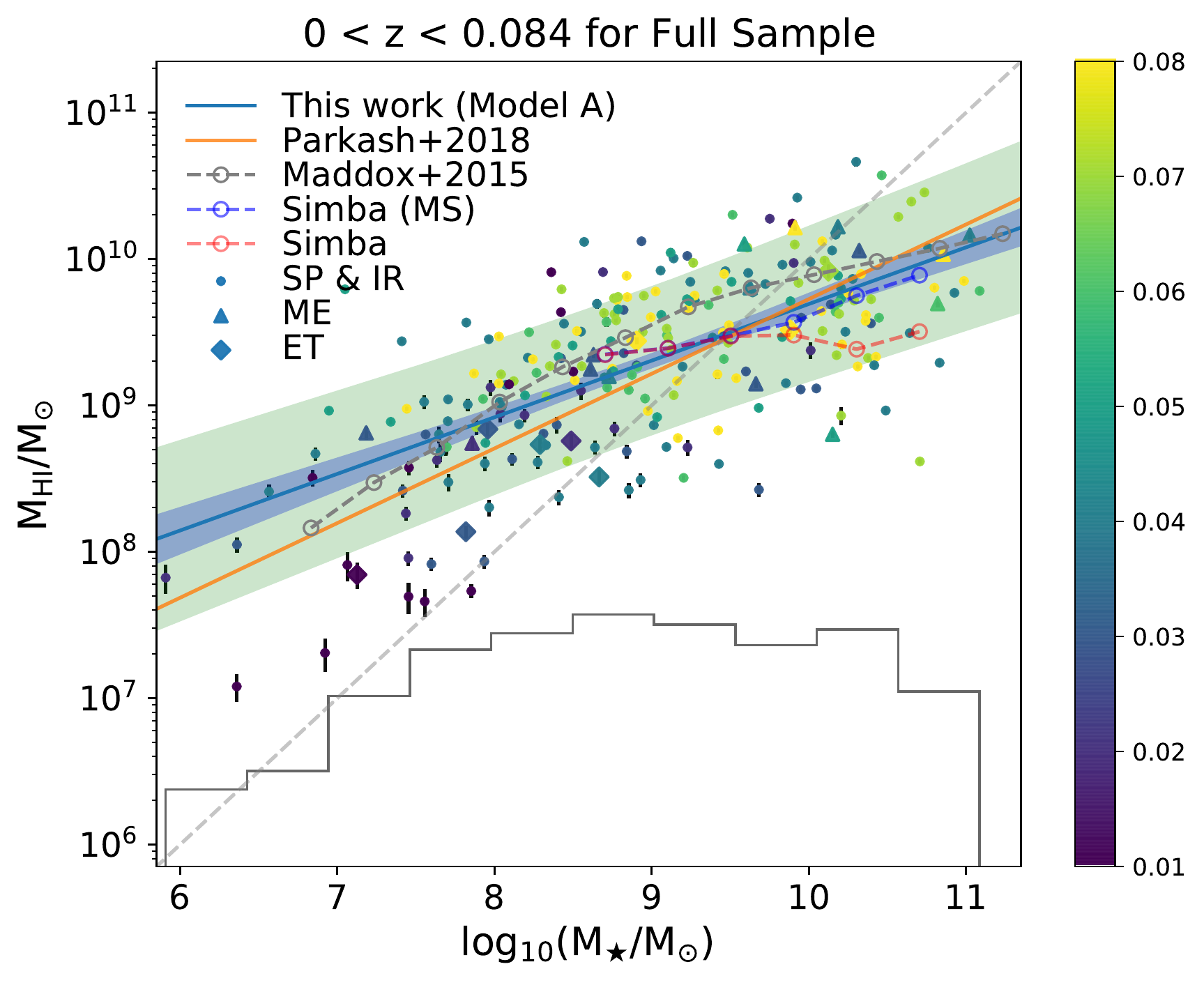}
    \includegraphics[width=0.95\columnwidth]{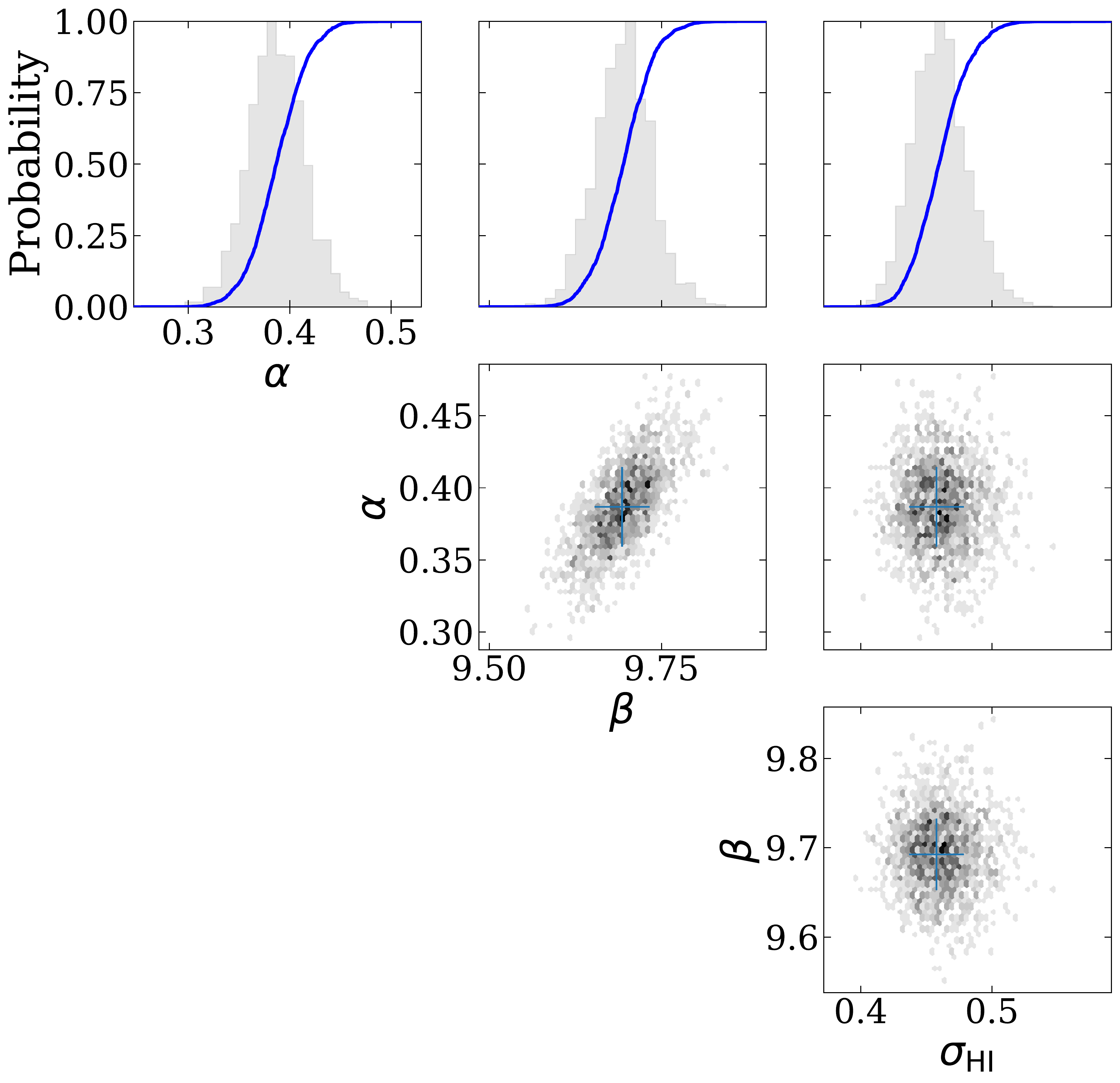}

  \end{subfigure}%
  \begin{subfigure}[s]{0.5\textwidth}
    \includegraphics[width=\columnwidth]{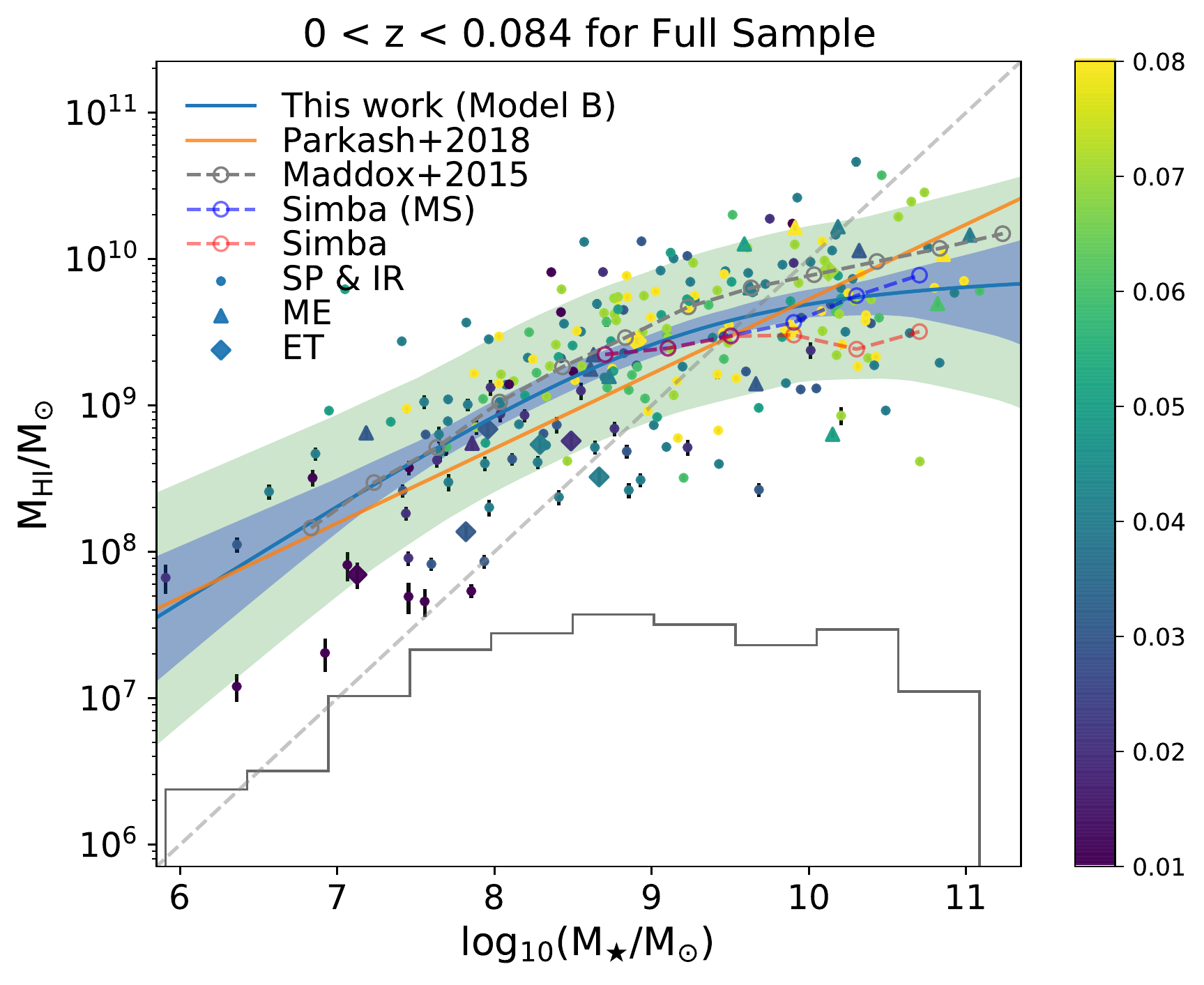}
    \includegraphics[width=0.95\columnwidth]{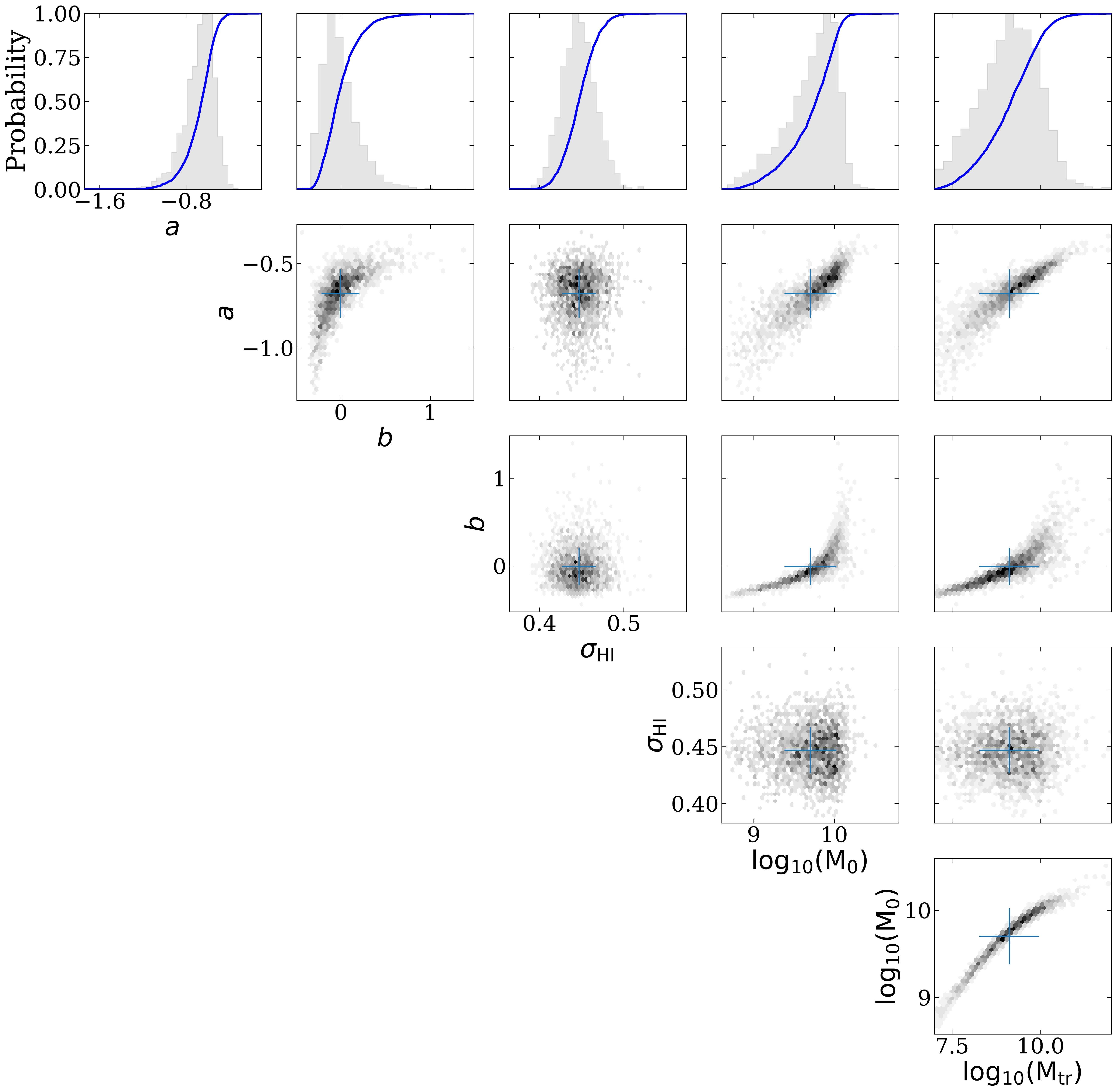}
  \end{subfigure}%
  \caption{Observed $M_{\rm HI}-M_{\star}$ relation (top) and its posterior parameters (bottom) with the best fitting Model A (left) and B (right) at  $0<z<0.084$ for the full \ha selected sample from MIGHTEE-\ha. Top row: The dots are spiral and irregular galaxies while triangles and diamonds  correspond to mergers and elliptical galaxies. They are colour-coded by redshift. The blue shaded areas are the statistical uncertainties, and the green ones are the intrinsic scatters added to the statistical uncertainties. The orange line is the \ha-selected sample from \protect\cite{Parkash2018}. The grey dashed circles are the measurements from ALFALFA \citep{Maddox2015}. The blue dashed circles are the main sequence galaxies (MS) from the SIMBA simulation \citep{Dav__2019} while the red ones are their full \ha samples. The diagonal light grey dashed line is the one-to-one relation. The black line at the bottom of the top panels indicates the normalised distribution of stellar mass. Bottom row: The grey histograms are the (1 or 2\,dimensional) marginal posterior probabilities. The blue curves are the cumulative distributions. The blue crosses in the 2 dimensional posteriors are the set of parameters with the maximum likelihood, and the 1$\sigma$ error bars error are estimated in the 1 dimensional marginal posterior space.}
  \label{fig:mstar-marg-full}
\end{figure*}

In this section, we investigate the relation between H{\sc i} mass and stellar mass based on our H{\sc i}-selected sample with the Bayesian method outlined in Section~\ref{sec:method} and mock samples. In Section~\ref{sec:upper_envelope}, we first use the full sample to maximise the baseline in stellar mass and H{\sc i} mass for modelling the upper envelope, and then consider the morphologically classified spiral galaxies as a separate population in order to compare with previous studies. In Section~\ref{sec:underlying}, we create mock samples to model the underlying $M_{\rm HI}-M_{\star}$ relation with two main galaxy populations based on the measured upper envelope of the $M_{\rm HI}-M_{\star}$ relation from the MIGHTEE-\ha catalogue.

\subsection{Modelling the upper envelope of the $M_{\rm HI}-M_{\star}$ relation}
\label{sec:upper_envelope}

\subsubsection{The whole sample}
\label{sec:hi-selected}
We show the \ha and stellar mass distribution for the complete \ha-selected sample in the top panels of Figure~\ref{fig:mstar-marg-full}. The best fitting lines for our linear Model A and non-linear Model B are shown as the blue lines in the left and right panels, respectively.  The 1$\sigma$ statistical scatter, predominantly due to the \ha flux uncertainties and our limited sample size are denoted by blue shaded areas, while the total (statistical plus intrinsic) scatter in  the \ha mass distribution around the stellar mass are shown by green shaded regions. We find that the non-linear model is decisively preferred over the linear model with a Bayes factor of $\Delta\ln(\mathcal{Z_B}) = 6.16^{+0.07}_{-0.07}$ for the full sample at $0<z<0.84$, and list the best fitting parameters in Table~\ref{tab:post-hi-AB}.

The agreement between the full MIGHTEE-H{\sc i} sample and the spectroscopic ALFALFA-SDSS galaxy sample of \cite{Maddox2015} is excellent for our non-linear Model B, with most parts of the \cite{Maddox2015} relation (denoted by grey dashed line) falling within the 1$\sigma$ statistical uncertainties of our data (blue shaded area). Compared to ALFALFA, the deficit at the high \ha mass (M$_{\rm HI} \gtrsim 10^{10}$\,M$_{\odot}$) end seen in both panels suggests we detected fewer \ha galaxies at these masses in the MIGHTEE-H{\sc i} Early Science data. This is due to the limited volume surveyed thus far which precluded us from finding the rarer, high H{\sc i}-mass systems in the current area, and we will require the full MIGHTEE survey, where the survey volume will reach 20 deg$^2$, to fully explore this region. Our results are in excellent agreement with the SIMBA simulation \citep{Dav__2019}, where we include the main sequence galaxies (MS) defined as specific SFR (sSFR) $> 10^{-1.8+0.3z}$ Gyr$^{-1}$ on top of the \ha selection that is identical to the MIGHTEE-\ha selection (blue dashed circles). This sSFR selection cuts out many red galaxies, and the  H{\sc i}-selected SIMBA sample shows no obvious systematic difference in typical sSFR compared to the MIGHTEE H{\sc i}-selected sample. However, we do find a deviation of the \ha masses between the H{\sc i}-selected MIGHTEE sample and the H{\sc i}-selected SIMBA sample without excluding red galaxies (red dashed circles). This deviation suggests that SIMBA overestimates the amount of \ha gas in the massive dead red galaxies as these galaxies seem to have a moderate amount of \ha gas and would have been detected by MIGHTEE-\ha, thus weighting down the average \ha mass at the massive end if they were present. The statistical significance of this difference is however quite low, and would need to be investigated with a larger sample. Given that the \ha column density is typically very low in red galaxies, another possibility is that our MIGHTEE-\ha Early Science observation could miss some of those objects.

Compared to the measurement from \cite{Maddox2015}, Model A would suggest an excess of \ha-rich galaxies at the low-mass end, but this is due to the model being a poor description of the data and this excess disappears when we use a more flexible non-linear model to fit for the data (in the right panel). The low-number statistics also plays a role at the low mass end as the statistical error, indicated by the blue area, increases. However, we note that the intrinsic scatter in the relation still dominates. The global difference between our linear Model A and the binned median \ha masses in \cite{Maddox2015} also suggests the limitation of a simple linear modelling, due to the complex nature of the upper bound in the $M_{\rm HI}-M_{\star}$ relation. Nonetheless, Model A is consistent with the \ha-selected sample in \cite{Parkash2018} at the high-mass end but a higher detection of rate of \ha galaxies at the low-mass end suggests that the \ha-selected sample in \cite{Parkash2018} is less complete at $M_{\rm HI} < 10^9M_\odot$.

For Model B, we find a transition stellar mass of $\log_{10}(M_\star/M_\odot)=9.15\pm0.87$, which breaks the upper envelope of the $M_{\rm HI}-M_{\star}$ relation into two regions. At the high mass end, the measured slope (indicated by the parameter $b$, is much flatter than that at the low mass end  (indicated by the parameter $a$). This finding is consistent with the steeper slope of Model A measured from the full \ha sample with respect to the spiral-only galaxies which tend to be massive systems (see also Section~\ref{sec:spirals}), and is also in line with the upper envelope of \ha mass fraction found by \cite{Maddox2015}. Thus, we confirm that the \ha gas fraction decreases as a function of stellar mass at $M_\star \gtrsim 10^9 M_{\odot}$ with the MIGHTEE-\ha Early Science data. This trend is similar to the galaxy main sequence, where the SFR-$M_\star$ relation is linear up to a critical mass of $\sim 3\times10^{10} M_\odot$, and then flattens out towards higher masses \citep[e.g.][]{Erfanianfar_2015}. It is also interesting to note that a similar curvature has been suggested for the baryonic specific angular momentum-baryonic mass relation with the slope change occurring at $\sim 10^9M_\odot$ \citep[e.g.][]{Kurapati_2018, Kurapati_2021}, albeit whether this break is real is still debated \citep{Mancera_Pi_a_2021,Mancera2021}.

The triangular and diamond symbols in Figure~\ref{fig:mstar-marg-full} denote mergers and elliptical galaxies respectively. We see the elliptical galaxies predominantly lie below the model fits, while the mergers are randomly distributed around the best-fit models. Thus, it shows that a lower fraction of \ha gas is detected in ellipticals compared to other types of galaxies from the \ha-selected sample as we might expect. However, we do not draw strong conclusions about this given their small number in our sample.

We note that our \ha-selected sample is flux-limited, and thus exhibits a selection bias against galaxies with low \ha masses, which becomes more severe going to higher redshift as seen from the colour-coded symbols. We return to this in Section \ref{sec:underlying}.

\begin{figure}
  \centering
  \includegraphics[width=\columnwidth]{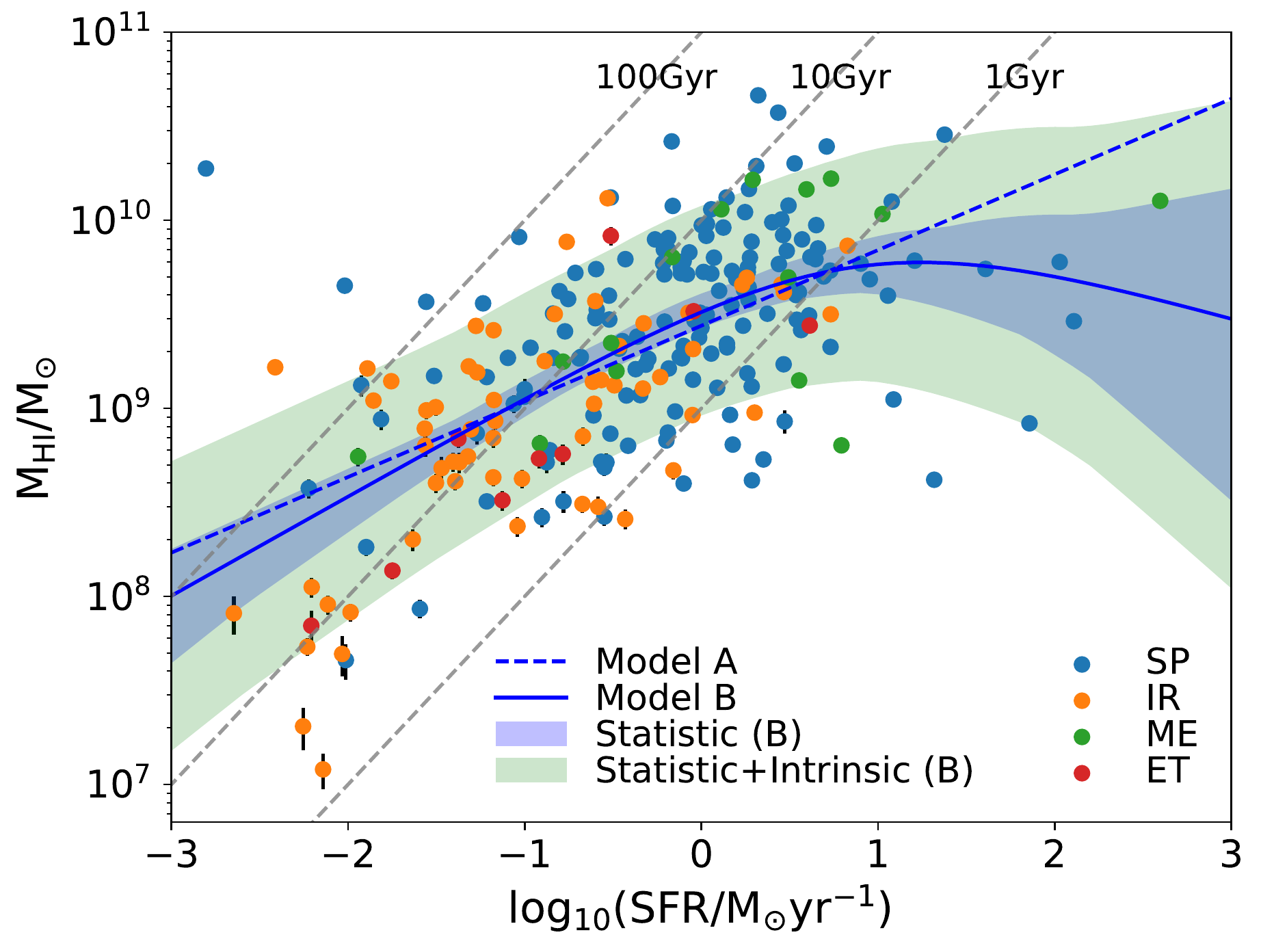}
  \caption{$M_{\rm HI}$ as a function of the SFR for the full \ha-selected sample from the MIGHTEE-\ha catalogue at $0<z<0.084$. Measurements in dots are colour-coded by their morphologies. The dashed and solid blue lines are the best fitting of Models A and B, respectively. The blue and green shaded areas are the statistical uncertainties and intrinsic scatters added to the statistical uncertainties for the Model B. The dashed grey lines are the time scales for depletion of the \ha gas, defined as $M_{\rm HI}$/SFR.}
  \label{fig:SFR-MHI}
\end{figure}

In Figure~\ref{fig:SFR-MHI}, we show the \ha mass as a function of SFR. We first replace the stellar mass with SFR in our models, then fit the Model A (dashed blue line) as $\langle\log_{10}(M_{\rm HI})\rangle = 0.4\log_{10}(\rm SFR)+9.44$ and the Model B (solid blue line), and observe a moderate flattening feature at the high SFR end with a measured transition SFR of $\log_{10}(\rm SFR/M_{\odot} yr^{-1})$ = 0.79$\pm0.53$, over which the statistical uncertainties are large due to only a few highest-SFR/bursty  spirals. We also find that the majority of our \ha-selected galaxies are able to support their star formation activity given a sufficient \ha fuel supply, with the \ha depletion times, $M_{\rm HI}$/SFR, varying in the range of 1Gyr to 100Gyr. This suggests that the correlation between SFR and the \ha mass is consistent with being almost linear across the entire \ha mass range on the logarithmic scale, and the shortage of \ha gas is likely ultimately responsible for the decreasing star formation rate towards the higher stellar masses, although we notice a slightly larger intrinsic scatter of $\sim0.48$ dex for this relation compared to the $\sim0.44$ dex for the $M_{\rm HI}-M_{\star}$ relation. The lower turnover mass of $\log_{10}(M_\star$/$M_{\odot}) = 9.15\pm0.87$ for the $M_{\rm HI}-M_{\star}$ relation against $\log_{10}(M_\star$/$M_{\odot}) \sim 10$ for the SFR-$M_\star$ relation signifies that the loss of \ha gas supply at high masses may not immediately reflected on the quenching of star formation, albeit with large statistical uncertainties.

\begin{figure*}
  \centering
  \begin{subfigure}[s]{0.5\textwidth}
    \includegraphics[width=\columnwidth]{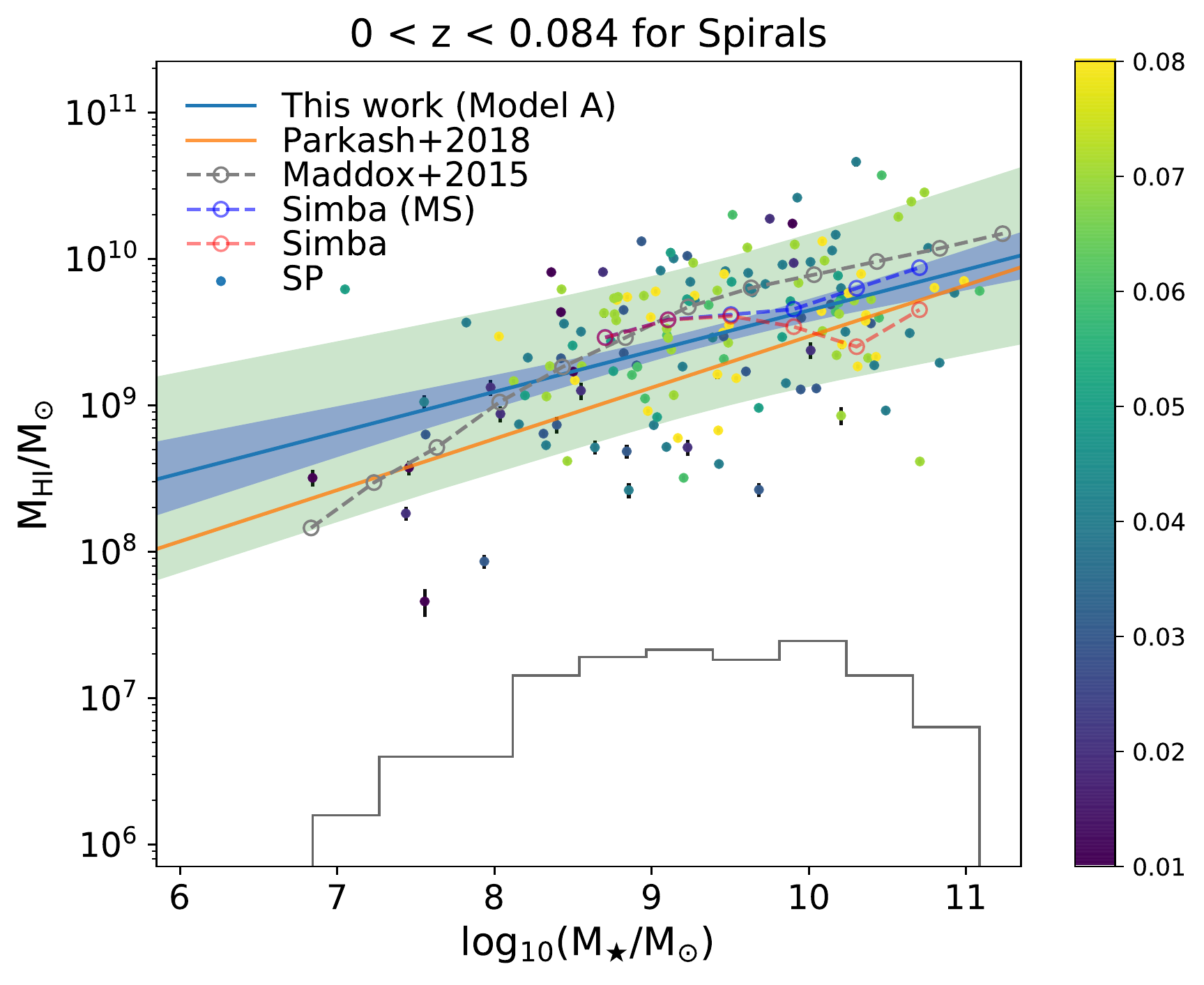}
    \includegraphics[width=0.95\columnwidth]{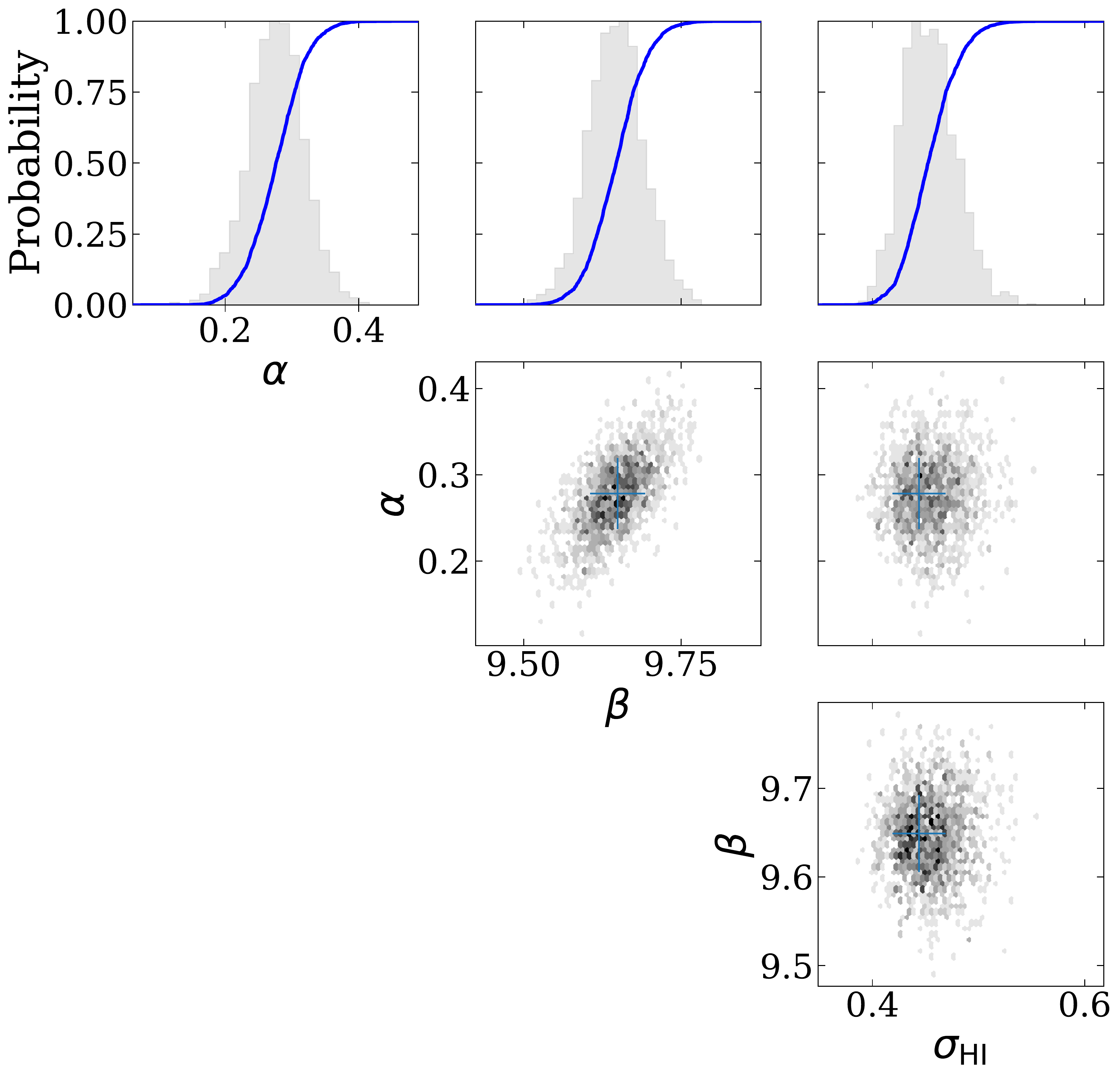}
  \end{subfigure}%
  \hfill
  \begin{subfigure}[s]{0.5\textwidth}
    \includegraphics[width=\columnwidth]{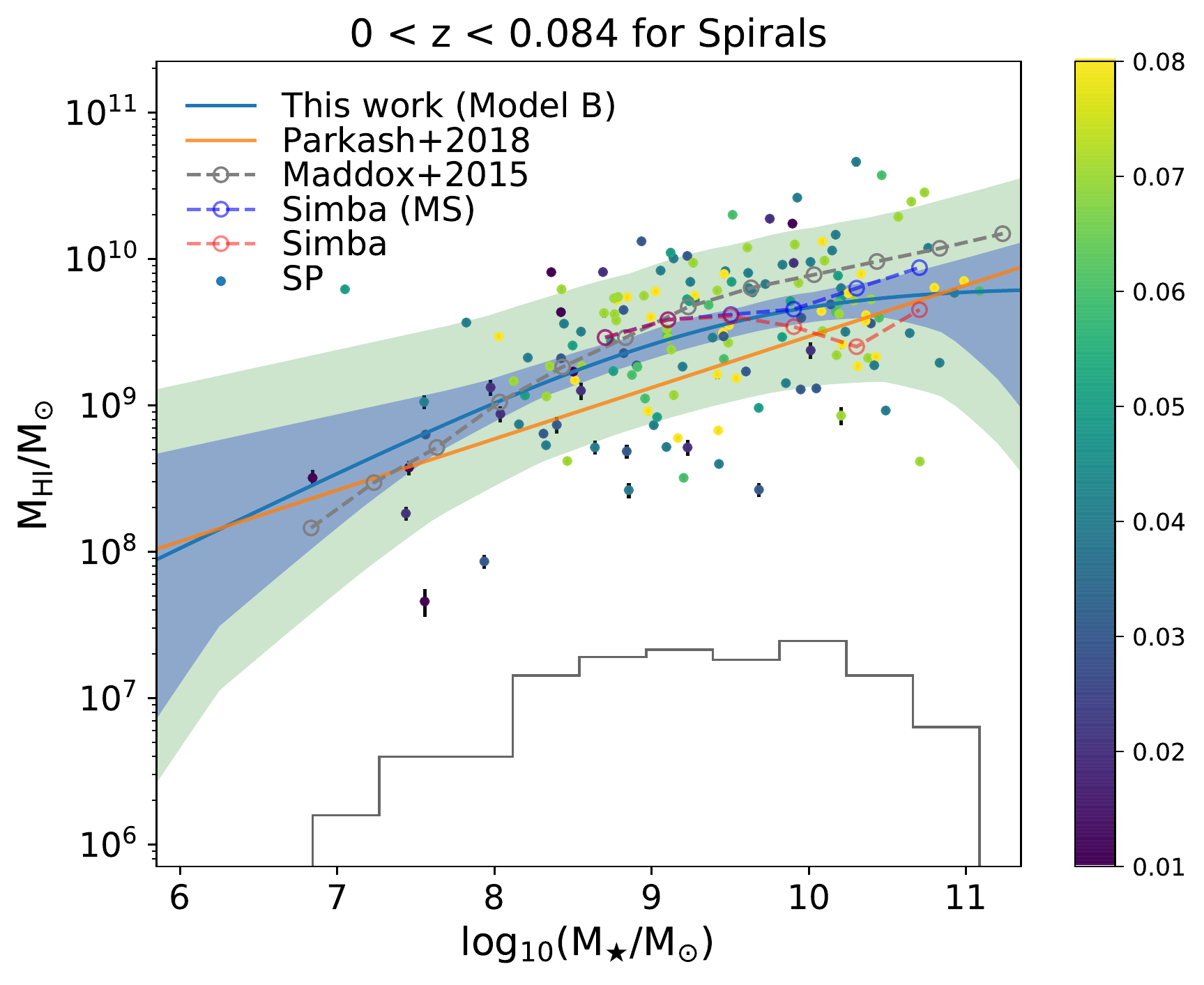}
    \includegraphics[width=0.95\columnwidth]{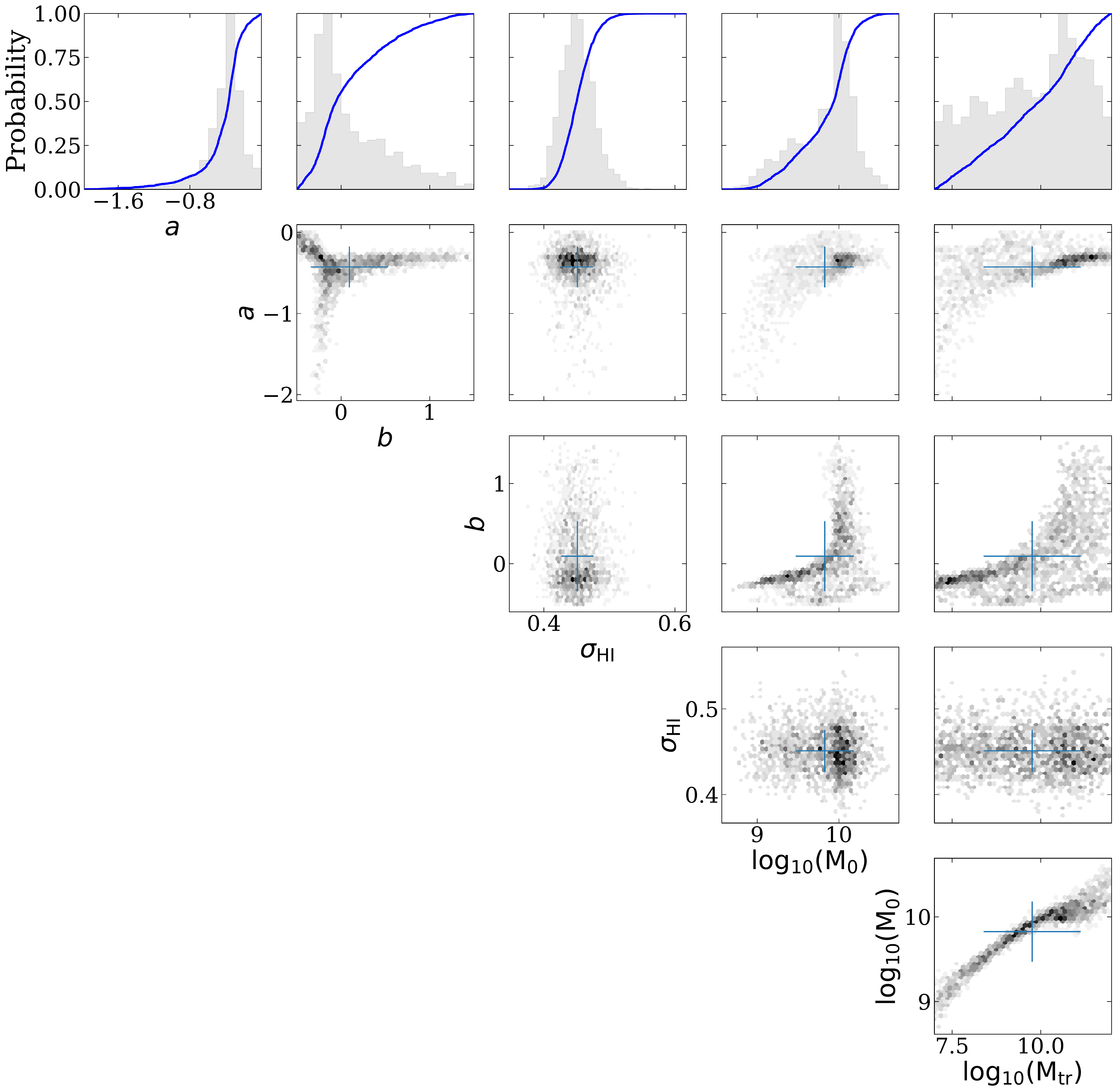}
  \end{subfigure}%
  \caption{Observed $M_{\rm HI}-M_{\star}$ relation  (top) and its posterior parameters (bottom) for spiral galaxies with the best fitting Model A (left) and B (right) at  $0<z<0.084$ from the MIGHTEE-\ha catalogue. Top row: The dots are spiral galaxies colour-coded by redshift. The blue shaded areas are the statistical uncertainties, and the green ones are the intrinsic scatters added to the statistical uncertainties. The orange line is the stellar mass-selected spirals from \protect\cite{Parkash2018}. The grey dashed circles are the measurements from ALFALFA \citep{Maddox2015}. Bottom row: The grey histograms are the (1 or 2\,dimensional) marginal posterior probabilities. The blue curves are the cumulative distributions. The blue crosses in the 2 dimensional posteriors are the set of parameters with the maximum likelihood, and the 1$\sigma$ error bars are estimated in the 1 dimensional marginal posterior space.}
  \label{fig:mstar-marg-sp}
\end{figure*}

\subsubsection{Spiral galaxies}
\label{sec:spirals}

In this section, we consider the population of morphologically classified spiral galaxies. In Figure~\ref{fig:mstar-marg-sp}, we show the observed $M_{\rm HI}-M_{\star}$ relation and the posterior parameters for the \ha-selected spiral galaxies from the MIGHTEE-\ha catalogue at $0<z<0.084$, with our best fitting Models A and B in the left and right top panels, respectively. Based on the best fits in Table~\ref{tab:post-hi-AB} and the returned Bayesian evidences, we see that the data are much less in favour of the non-linear model over the linear one with $\Delta\ln(\mathcal{Z_B}) = 1.44\pm0.04$, which is significant but not decisive or strong. We also find that the posterior distributions of Model B for spirals are not as well-converged as for the full sample in Figure~\ref{fig:mstar-marg-full}.

For both models, we find a systematically higher detection of \ha gas than what \cite{Parkash2018} found at $M_{\star}\gtrsim 10^9 M_\odot$ from a stellar mass-selected spirals. This is likely to be the result of different selection effects, with the \cite{Parkash2018} spiral galaxy sample being $M_\star$-selected and the MIGHTEE-\ha sample being \ha-selected. The latter tends to be populated by higher \ha mass objects at any given stellar mass. It implies that there still exists a significant fraction of \ha-poor spiral galaxies to be picked up by a deeper \ha survey. We measure an intrinsic scatter of $0.44\pm0.03$ dex for both models, which is roughly consistent with the 0.4 dex obtained from the stellar mass-selected spirals in \cite{Parkash2018}.

To compare with the SIMBA spirals, we select galaxies in SIMBA with fraction of kinetic energy ($\kappa_{\rm rot}$) > 0.7 \citep{Sales_2012}, and denote their median \ha masses against the stellar masses as red dashed circles. We then use the same criterion of sSFR $> 10^{-1.8+0.3z}$ Gyr$^{-1}$ to exclude the red spiral galaxies, and show their $M_{\rm HI}-M_{\star}$ relation as blue dashed circles. Overall, we find good agreement between MIGHTEE-\ha and SIMBA MS spirals for the $M_{\rm HI}-M_{\star}$ relation, and notice a lower detection of the average \ha mass for the whole SIMBA spiral sample at the most massive end. This trend is similar to what we found in Section~\ref{sec:hi-selected} for the full MIGHTEE-\ha sample, and indicates that there are probably too many red spiral galaxies that have non-negligible amount of \ha gas in SIMBA.  

The best fitting transition stellar mass for our Model B for the \ha-selected spirals is $\log_{10}(M_\star$/$M_{\odot})$ = $9.52\pm1.56$, which is higher than the $\log_{10}(M_\star$/$M_{\odot})$ = $9.15\pm0.87$ for the full sample, but has much larger statistical uncertainties due to reduced number of sources and a narrower range of $M_\star$. This trend roughly corresponds to the difference of the best fitting slopes using Model A between these two samples, where the spirals have an obvious shallower slope of 0.278 compared to the 0.387 for the full sample, although the corresponding intrinsic scatters are similar.

The distinction of our best fitting models between spirals and the full \ha-selected sample is a strong indication of very different gas processes between the spiral and lower-mass irregular galaxies, since the irregulars dominate over other types of galaxies except the spirals in our catalogue. Indeed, given the stability model for disk galaxies in \cite{Obreschkow_2016}, the halo spin parameter for the spiral galaxies can limit the maximum \ha gas supply, and there seems to be no such a limitation for the lower-mass galaxies as their disks become unstable when the velocity dispersion reaches similar to rotation velocity. However, we cannot distinguish whether the different slopes are due to galaxy mass or morphology with the current sample, and this should be better explored with the full MIGHTEE survey. Stacking on the spiral galaxies to lower stellar mass and other types of galaxies at higher stellar mass will also help to further clarify this difference.

\subsection{Modelling the underlying $M_{\rm HI}-M_{\star}$ relation}
\label{sec:underlying}

\begin{figure*}
  \centering
  \begin{subfigure}[b]{0.5\textwidth}
    \includegraphics[width=0.95\columnwidth]{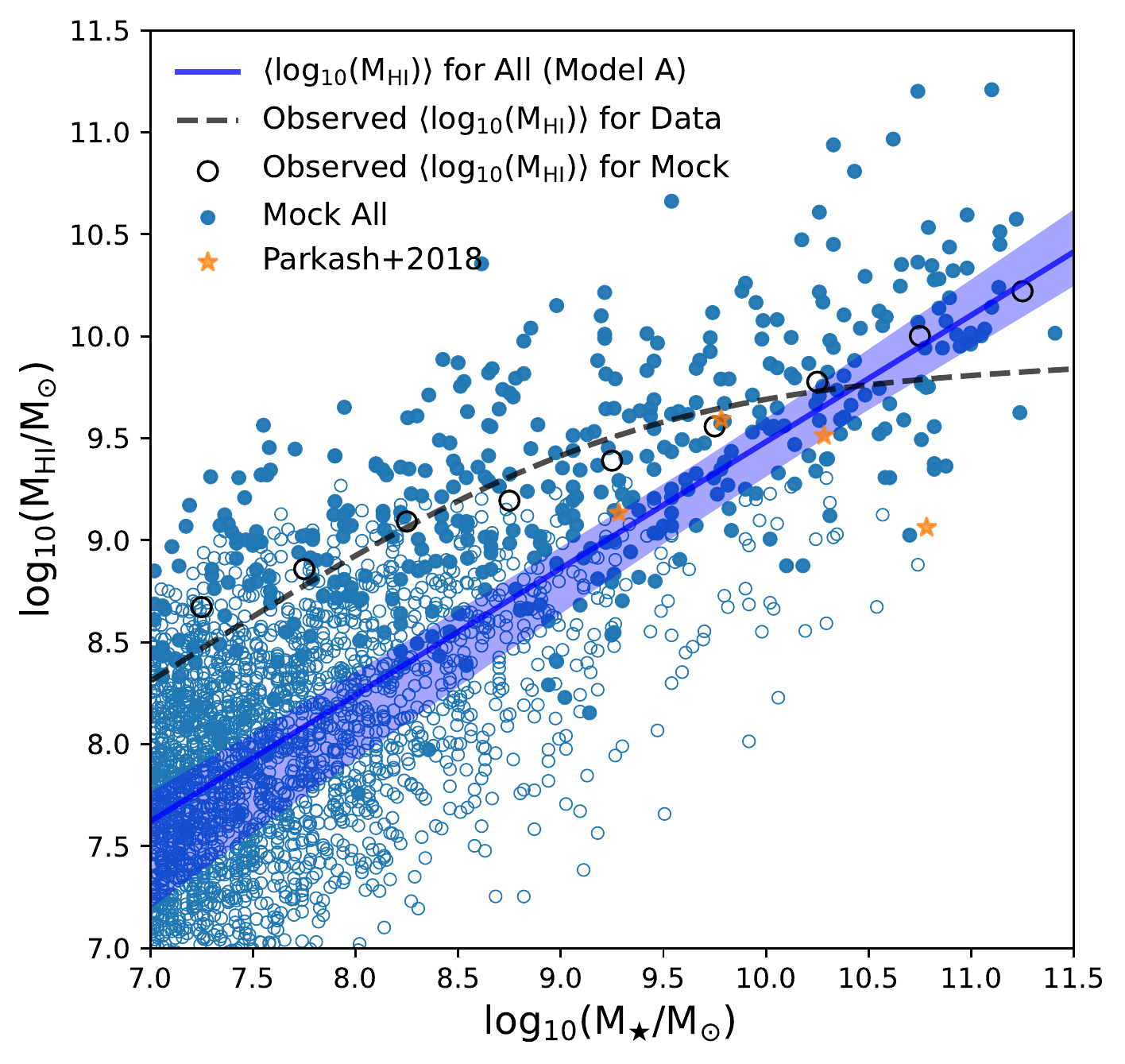}
    \includegraphics[width=0.95\columnwidth]{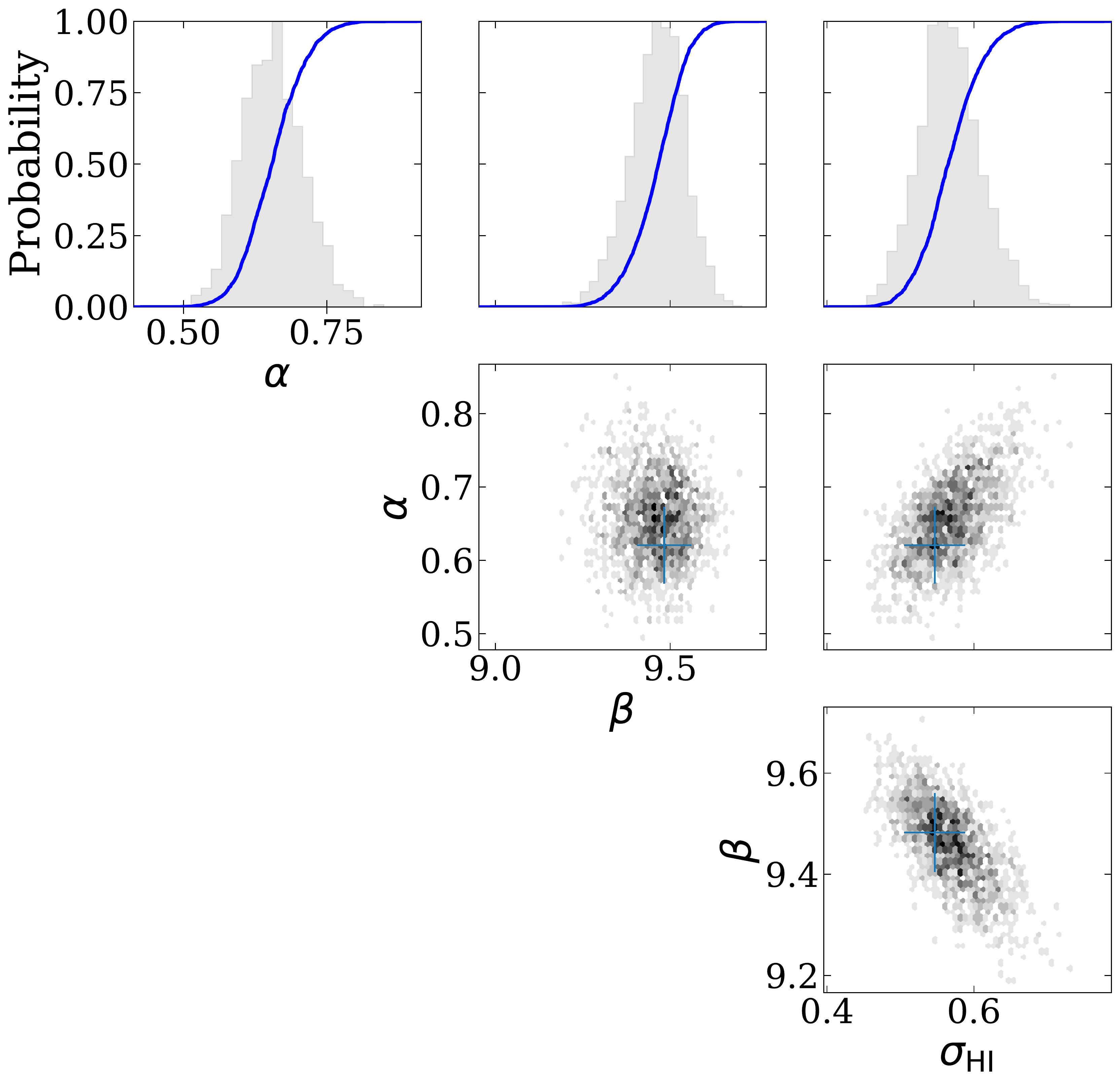}
  \end{subfigure}%
  \hfill
  \begin{subfigure}[b]{0.5\textwidth}
    \includegraphics[width=0.95\columnwidth]{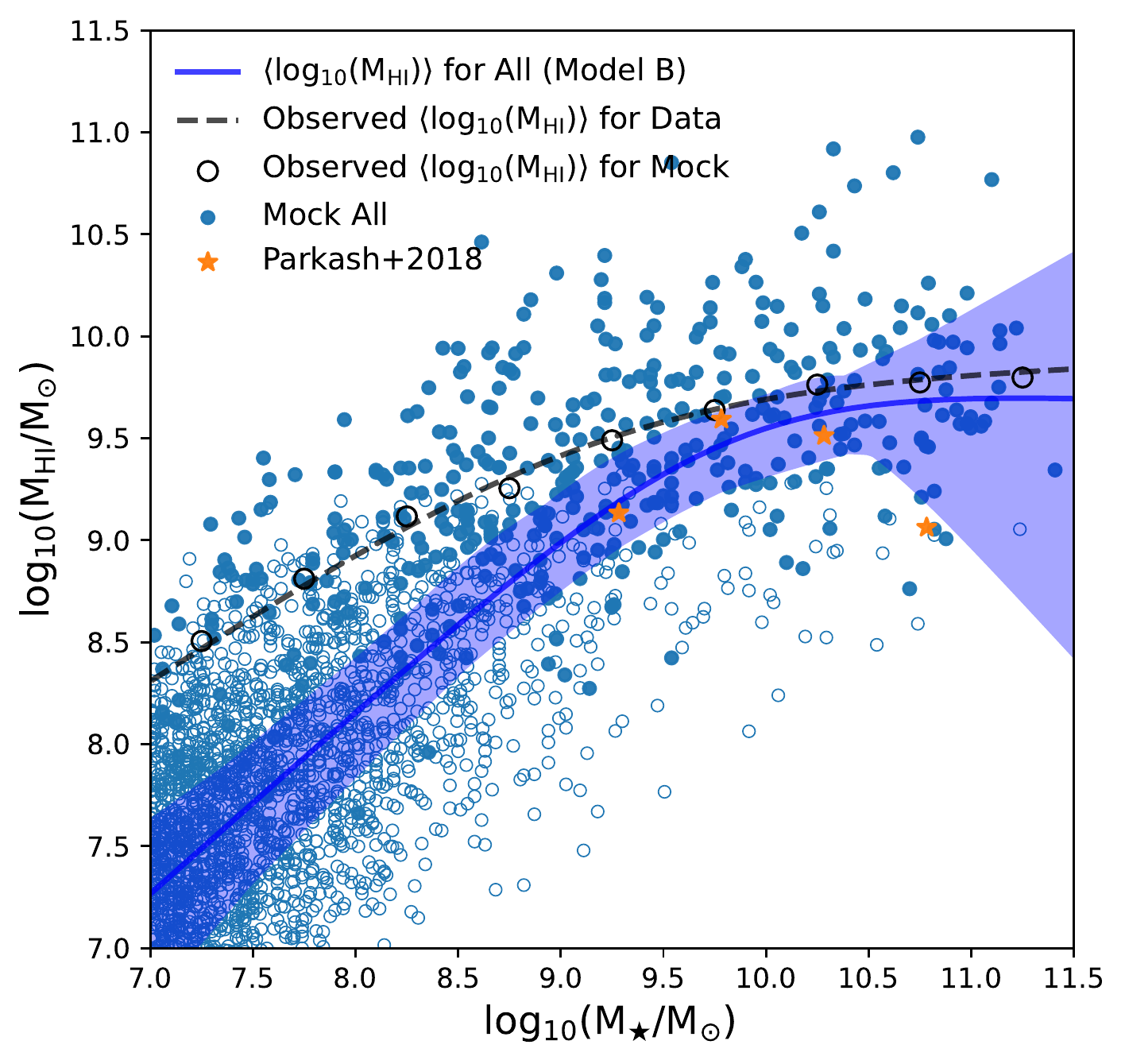}
    \includegraphics[width=0.94\columnwidth]{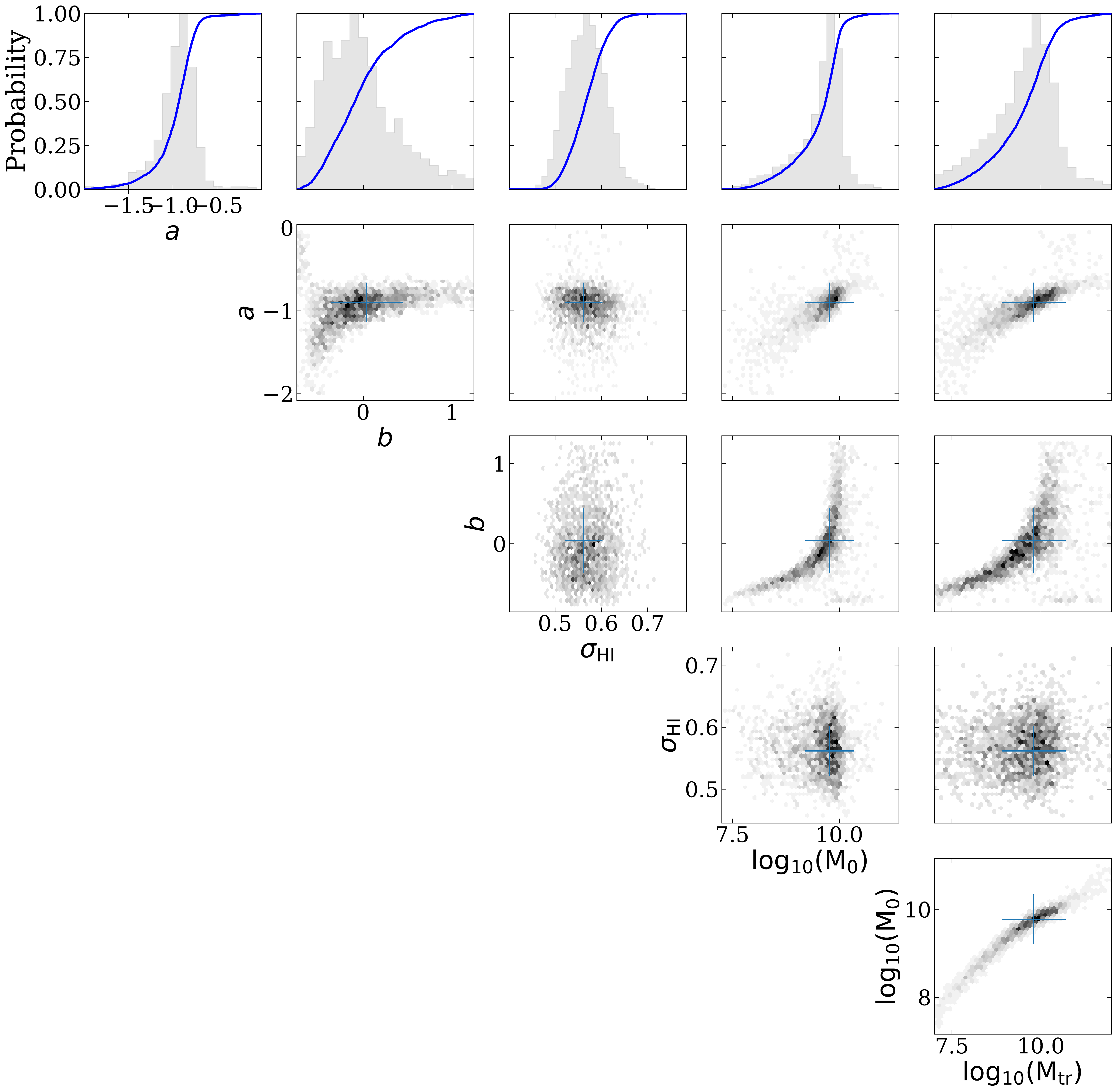}
  \end{subfigure}%
  \caption{Top row: $M_{\rm HI}$ as a function of the stellar mass for the whole mock MIGHTEE-\ha sample at $0<z<0.084$, with Model A (left) and B (right) fitted for the underlying $M_{\rm HI}-M_{\star}$ relation. The blue dots with the filled and empty ones are the detections and non-detections. The dashed black lines in the left and right panels are the observed $\langle\log_{10}(M_{\rm HI})\rangle$ from Model B for our MIGHTEE-\ha observation while the black circles are "observed" $\langle\log_{10}(M_{\rm HI})\rangle$ for our mock sample. The solid blue lines are the best fitting of Models A and B in the left and right panels for the whole underlying sample with the blue shaded areas being the 1$\sigma$ statistical uncertainties. The orange stars are the stellar mass-selected sample from \protect\cite{Parkash2018}. The blue dots are down-sampled for presentation. Bottom row: The grey histograms are the (1 or 2\,dimensional) marginal posterior probabilities. The blue curves are the cumulative distributions. The blue crosses in the 2 dimensional posteriors are the set of parameters with the maximum likelihood, and the 1$\sigma$ error bars are estimated in the 1 dimensional marginal posterior space.}
  \label{fig:mstar-hi-D}
\end{figure*}

Thus far, we have discussed the $M_{\rm HI}-M_{\star}$ relation based on the \ha-selected sample from MIGHTEE-\ha Early Science data. The \ha selection essentially means that we are exploring the upper envelope in \ha mass and therefore constraining the maximum amount of \ha as a function of stellar mass. To fully assess how the \ha selection influences our results, in this section we create mock samples of all the galaxies above and below the MIGHTEE-\ha detection threshold with two main galaxy populations: late-type galaxies (LTGs) and early-type galaxies (ETGs). This also allows us to check how the limited volume and the flux-density limit selection of our sample may influence the results presented in the previous section when determining whether there is any evidence for a transition in the upper envelope of the stellar-mass to \ha-mass relation.

\subsubsection{The whole sample}
\label{sec:underlying_whole}

We create mock MIGHTEE-\ha galaxies with and without a break in the underlying $M_{\rm HI}-M_{\star}$ relation when considering the LTGs and ETGs as our whole sample, as shown in Figure~\ref{fig:mstar-hi-D}. We first employ the galaxy stellar mass function (GSMF) from \cite{Driver_2022} and generate galaxy samples across the redshift range of the MIGHTEE-\ha observations, but with 10 times the survey area to reduce the random error in this process. We then find the best fitting parameters of the underlying models for the $M_{\rm HI}-M_{\star}$ relation such that our mock \ha distribution matches our modelled upper envelope described by the average \ha mass $\langle\log_{10}(M_{\rm HI})\rangle$ and the global intrinsic scatter $\sigma_{\rm HI}$, above the \ha detection threshold in each MIGHTEE-\ha field. Considering the large mock survey area and the fact that no measurement uncertainty is introduced for mock data, we first bin the sample in $M_{\star}$ and  obtain the best fitting parameters by minimizing the difference between observed $\langle\log_{10}(M_{\rm HI})\rangle$ from our Model B for MIGHTEE-\ha (derived in Table~\ref{tab:post-hi-AB}) and for our mock sample as a function of $M_{\star}$, along with their $\sigma_{\rm HI}$, based on the same Bayesian framework. In general, the fitting for a 2-dimensional distribution to another one requires binning the data in the 2-dimensional space. However, our MIGHTEE-\ha sample size is relatively small and the distribution of $M_{\rm HI}$ against $M_{\star}$ can be well-described by a scaling relation and an associated scatter as shown in the previous section, therefore such a fitting in our case can be achieved by constraining the $\langle\log_{10}(M_{\rm HI})\rangle$ and $\sigma_{\rm HI}$, which are the key elements that delineate a 2-dimensional distribution in a simple way. We list the best fitting parameters in Table~\ref{tab:post-hi-C}.

In Figure~\ref{fig:mstar-hi-D}, we show the best fitting results in blue lines with our Models A and B in the left and right panels, respectively. The filled and empty symbols represent the "detections" and non-detections, respectively. Although it appears that both models can mimic a broken $M_{\rm HI}-M_{\star}$ relation for our MIGHTEE-\ha observation, the relative evidence between Models A and B for fitting the underlying $M_{\rm HI}-M_{\star}$ relation is $\ln(\mathcal{Z_B})$ - $\ln(\mathcal{Z_A}) = 3.9 \pm 0.3$, which strongly  implies a non-linear underlying $M_{\rm HI}-M_{\star}$ relation over the linear one. The black circles are the observed $\langle\log_{10}(M_{\rm HI})\rangle$ for our mock samples with the underlying Models A and B in the left and right panels, respectively. It is clear that the “observed” broken relation of the Model B mock sample agrees with the data (the dashed black line) better than the “observed” broken relation of the Model A mock sample especially at around the transition mass range. The non-linear model also demonstrates a better agreement than the linear model for the averaged \ha mass with \cite{Parkash2018}, based on their $M_{\star}-$selected sample.

We also perform a 2-dimensional Kolmogrov-Smirnov (KS) test\footnote{\url{https://github.com/syrte/ndtest}} \citep{Peacock1983, Fasano1987} to check consistency of distributions for the mock “detections” from our underlying models and the MIGHTEE-\ha detections. If the KS statistic is large, the p-value will be small, and it can be taken as evidence against the null hypothesis that the two distributions are identical. We accept a confidence level of 95$\%$, which means that we reject the null hypothesis if the p-value is less than 0.05. We find that the p-value is 0.129 for the mock detections from our Model B against our MIGHTEE-\ha data, and
the p-value is 0.007 for the mock detections from the underlying Model A. These p-values suggests that the non-linear underlying model is doing a better job in mocking the MIGHTEE-\ha detections than the linear one, which is consistent with what we see from the Bayesian evidences.

\subsubsection{Late-type galaxies}
\label{sec:underlying_ltgs}

\begin{figure*}
  \centering
  \begin{subfigure}[b]{0.5\textwidth}
    \includegraphics[width=\columnwidth]{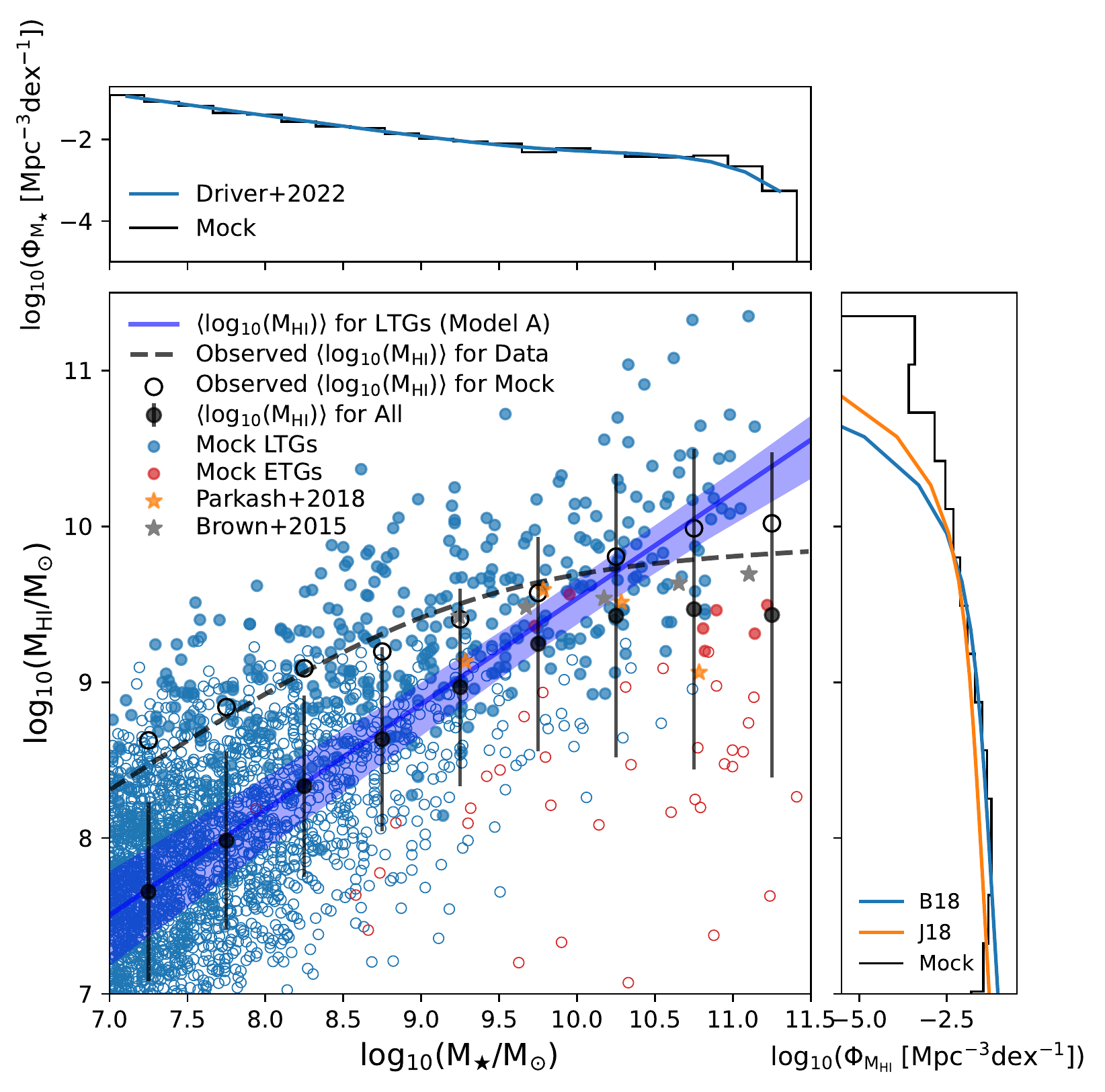}
    \includegraphics[width=0.95\columnwidth]{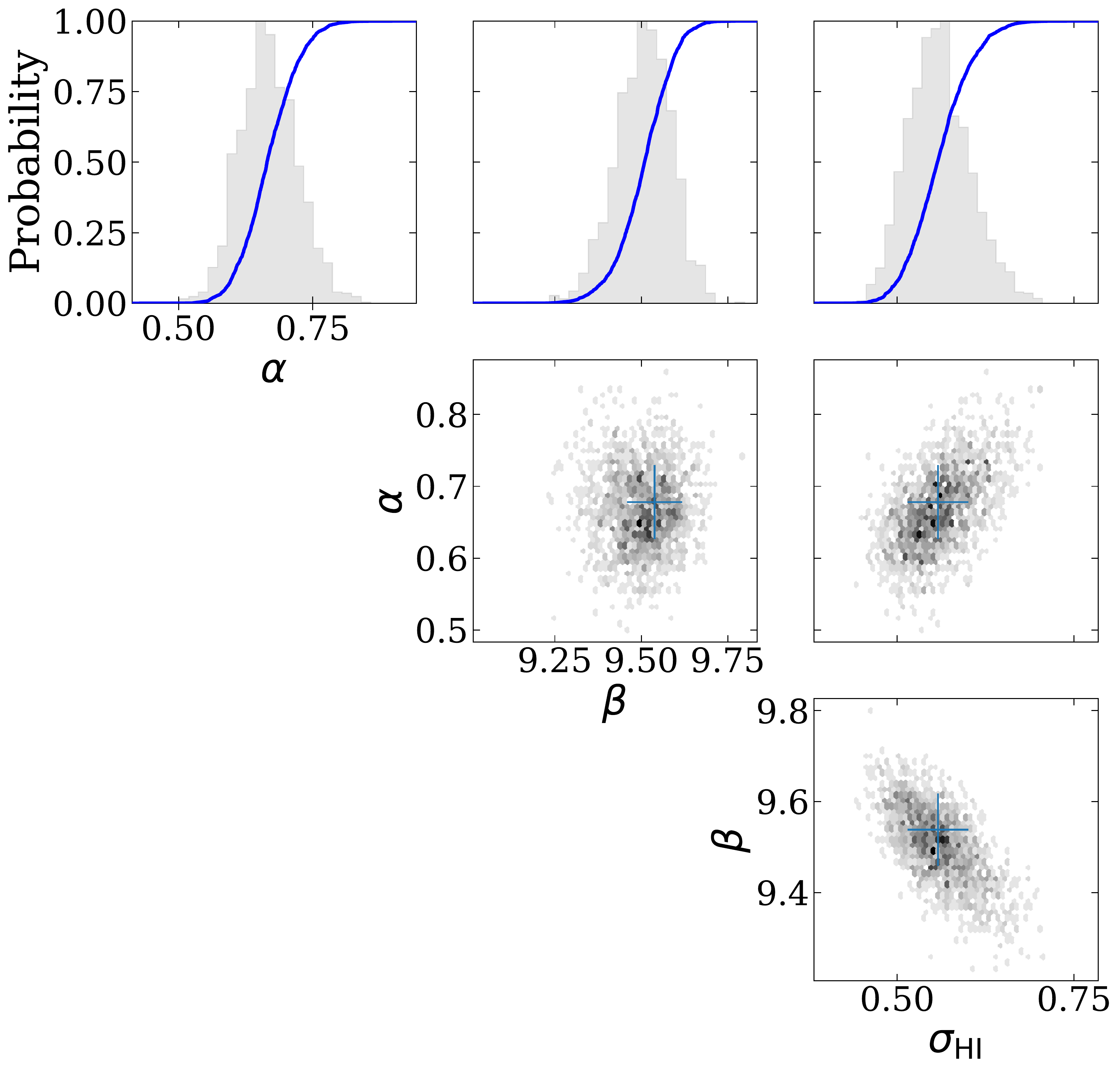}
  \end{subfigure}%
  \hfill
  \begin{subfigure}[b]{0.5\textwidth}
    \includegraphics[width=\columnwidth]{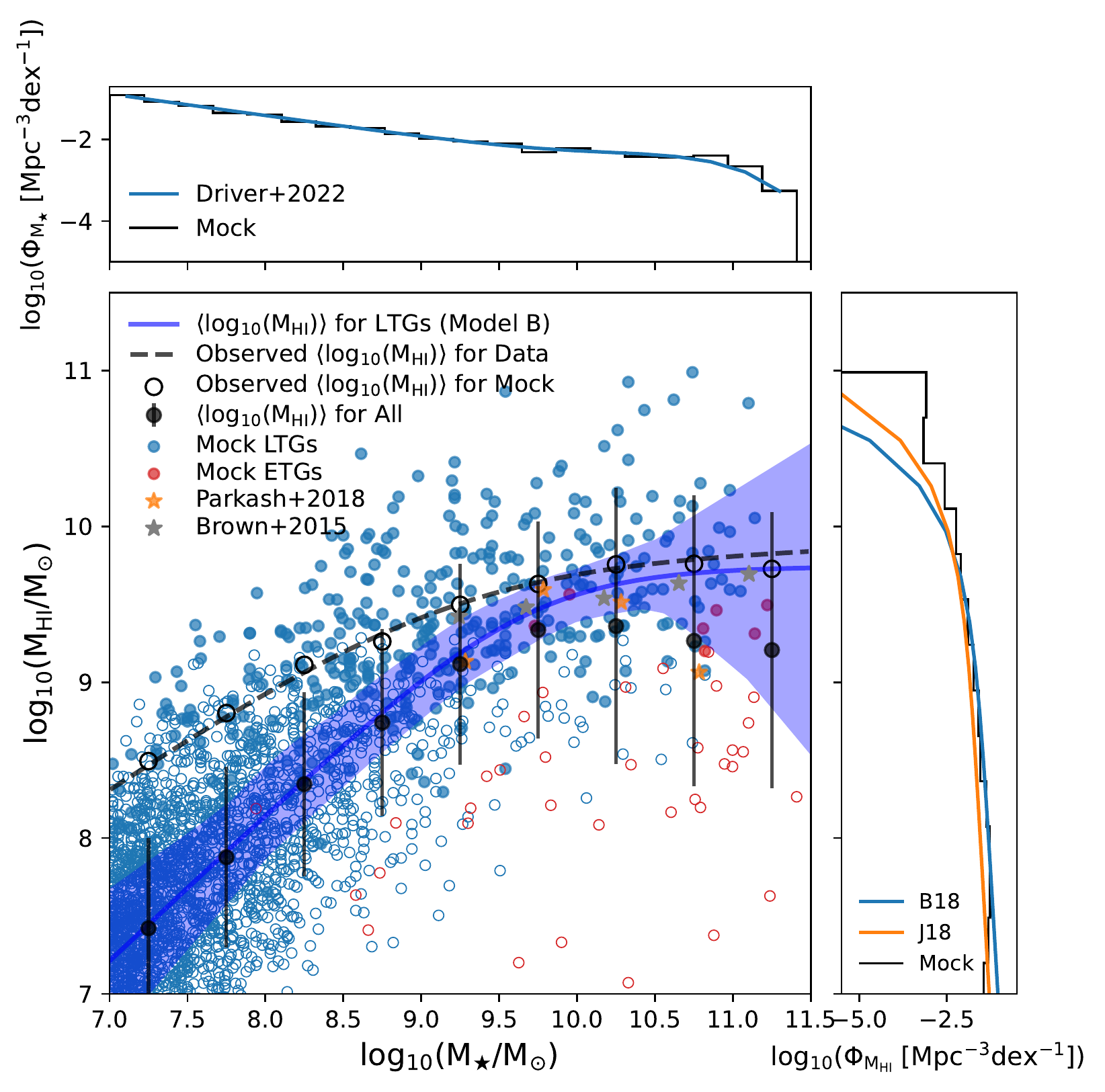}
    \includegraphics[width=0.93\columnwidth]{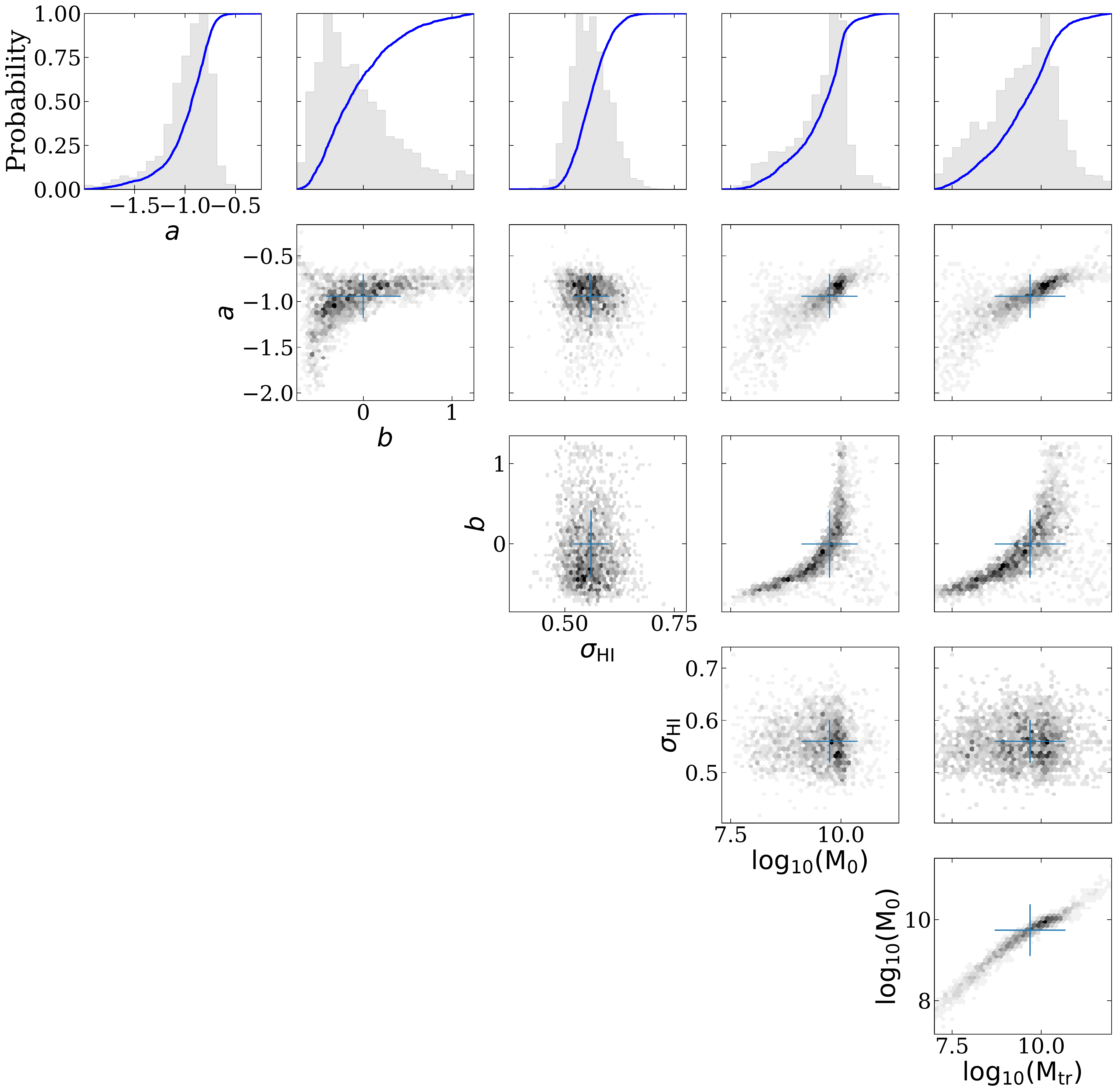}
  \end{subfigure}%
  
  \caption{Top row: $M_{\rm HI}$ as a function of the stellar mass for the mock MIGHTEE-\ha samples for LTGs and ETGs at $0<z<0.084$, with Model A (left) and B (right) fitted for the underlying $M_{\rm HI}-M_{\star}$ relation for LTGs. The blue and red dots are LTGs and ETGs, with the filled and empty ones being the detections and non-detections. The dashed black lines in the left and right panels are the observed $\langle\log_{10}(M_{\rm HI})\rangle$ from Model B for our MIGHTEE-\ha observation while the black circles are "observed" $\langle\log_{10}(M_{\rm HI})\rangle$ for our mock sample. The solid blue lines are the best fitting of Models A and B in the left and right panels for LTGs with the blue shaded areas being the 1$\sigma$ statistical uncertainties. The black dots are the average of $\log_{10}(M_{\rm HI})$ with the error bar being the intrinsic scatter for all samples at given stellar mass bins. The orange and grey stars are the stellar mass-selected samples from \protect\cite{Parkash2018} and \protect\cite{Brown_2015}. The upper panel shows the GSMF from \protect\cite{Driver_2022} that we use to mock our galaxy samples, while the bottom right panel shows the HIMF which is the result of marginalzing over the $M_\star$ axis. The HIMFs in \protect\cite{Jones_2018} for the ALFALFA survey and \protect\cite{Butcher_2018} for the NIBLES survey are shown in blue and orange lines. The color-coded dots are down-sampled for presentation. Bottom row: The grey histograms are the (1 or 2\,dimensional) marginal posterior probabilities. The blue curves are the cumulative distributions. The blue crosses in the 2 dimensional posteriors are the set of parameters with the maximum likelihood, and the 1$\sigma$ error bars are estimated in the 1 dimensional marginal posterior space.}
  \label{fig:mock_C}
\end{figure*}

\renewcommand{\arraystretch}{1.2}
\begin{table}
	\centering
	\caption{The best fitting parameters of the underlying $M_{\rm HI}-M_{\star}$ relation with Models A and B for our mock MIGHTEE-\ha samples at $0<z<0.84$. The values listed are those with the maximum likelihood from our fitting.}
	\label{tab:post-hi-C}
 	\begin{adjustbox}{max width=\columnwidth}
	\begin{tabular}{clrr}
	\hline
        \hline
	Model & Parameter  & The whole sample & Late-type galaxies \\
	\hline
          & $\alpha $         &0.621$\pm$0.053  & 0.678$\pm$0.052  \\
	  A   & $\beta$          &9.48$\pm$0.075    & 9.54$\pm$0.08 \\
	      & $\sigma_{\rm HI}$&0.55$\pm$0.04     & 0.56$\pm$0.04 \\
    \hline
         &  $a$                &-0.896$\pm$0.188       & -0.941$\pm$0.203 \\
          &  $b$                 &0.040$\pm$0.404      & -0.001$\pm$0.419 \\
	  B  & $\sigma_{\rm HI}$  &0.562$\pm$0.042       & 0.56$\pm$0.042 \\
	     & $\log_{10}(M_0)$     &9.77$\pm$0.51          & 9.74$\pm$0.62  \\
       & $\log_{10}(M_{\rm tr})$  &9.8$\pm$0.88         &   9.69$\pm$1.03 \\
    \hline
	\end{tabular}
	\end{adjustbox}
\end{table}

\renewcommand{\arraystretch}{1}
\begin{table}
	\centering
	\caption{Logarithmic average of $M_{\rm HI}$ and its intrinsic scatter as a function of stellar mass for the mock MIGHTEE-\ha LTGs and ETGs including detections and non-detections.}
	\label{tab:post-hi-mock}
	\begin{adjustbox}{max width=\textwidth}
	\begin{tabular}{c|cc}
	\hline
	\hline
         &\multicolumn{2}{c}{0.0$<$$z$$<$0.084}\\
           $\log_{10}(M_\star)$  & $\langle\log_{10}(M_{\rm HI})\rangle$ & $\sigma_{\rm HI}(M_\star)$ 
                \\ \hline 
                7.25 &  7.42 & 0.58 \\
                7.75 &  7.88 & 0.58 \\
                8.25 &  8.35 & 0.59 \\
                8.75 &   8.74 & 0.6 \\
                9.25 &  9.12 & 0.65 \\
                9.75 &   9.33 & 0.7 \\
                10.25 &  9.36 & 0.89 \\
                10.75 &  9.27 & 0.93 \\
                11.25 &  9.21 & 0.89 
                \\ \hline
	\end{tabular}
	\end{adjustbox}
\end{table}

We now divide the whole galaxy sample into LTGs and ETGs based on their known relative fractions as a function of stellar mass \citep{Rodr_guez_Puebla_2020}, and we assign the \ha masses with a double power law for ETGs \citep{Calette_2018}. The LTGs dominate the sample in our mass range of interest ($7 < \log_{10}(M_\star$/$M_{\odot}) < 11$), and we only consider tuning our Models A and B for LTGs in order to find the best fitting parameters that make our mock \ha distribution satisfy the observed $\langle\log_{10}(M_{\rm HI})\rangle$ and the global intrinsic scatter $\sigma_{\rm HI}$. We also list the best fitting parameters of the intrinsic $M_{\rm HI}-M_{\star}$ relation for LTGs in Table~\ref{tab:post-hi-C}.

In Figure~\ref{fig:mock_C}, we show the mock galaxy samples, along with the best fitting intrinsic $M_{\rm HI}-M_{\star}$ relation (solid blue line) for LTGs with our Models A and B in the left and right panels. The observed $M_{\rm HI}-M_{\star}$ relation with Model B from our MIGHTEE-\ha observation at $0<z<0.084$ are shown by dashed black lines in both panels. The blue and red dots are LTGs and ETGs, with the filled and empty symbols representing the "detections" and non-detections. The black dots are the average of $\log_{10}(M_{\rm HI})$ with the error bar being the intrinsic scatter for all samples at a given stellar mass bin (also shown in Table~\ref{tab:post-hi-mock}), and are in good agreement with the median of $\log_{10}(M_{\rm HI})$ for the stellar mass-selected sample in \cite{Parkash2018}. Compared to \cite{Brown_2015}  (grey stars), our measured average \ha masses appear to be systematically lower, which actually is not surprising as we are using the logarithmic average of \ha masses $\langle\log_{10}(M_{\rm HI})\rangle$ against the arithmetic average of \ha masses $\log_{10}(\langle M_{\rm HI}\rangle)$ in \cite{Brown_2015}. We find that the \ha selection lifts up the  "observed" $M_{\rm HI}-M_{\star}$ relation but the turnover feature on this relation persists to a large degree, albeit with a weaker break on the observed $M_{\rm HI}-M_{\star}$ relation. In other words, the \ha sample selection has a stronger impact on the lower mass end for the $M_{\rm HI}-M_{\star}$ relation as the dwarf galaxies are most sensitive to our detection floor, and hence the underlying average \ha masses have a sharper bend than the observed average \ha masses as a function of the stellar mass. Although the ETGs can downweight the average \ha mass at the high mass end, it is obvious that the LTGs alone show an intrinsic turnover on the $M_{\rm HI}-M_{\star}$ relation as indicated by the solid blue line.

By marginalising over the stellar mass in Figure~\ref{fig:mock_C}, we can also obtain a best fitting \ha mass function (HIMF) shown by the black line in the bottom right panel. The HIMF constructed by our approach is in good agreement with \cite{Jones_2018} and \cite{Butcher_2018} across a wide range of \ha masses, except for the highest mass end ($\log_{10}(M_{\rm HI}$/$M_{\odot}) \gtrsim10$) due to the unsophisticated modelling (e.g. the assumption of a symmetric \ha distribution at a given stellar mass for each population) for the bivariate HI-stellar mass distribution. We refer readers to a more detailed approach to measuring the first MeerKAT HIMF by \cite{ponomareva2023mighteehi}. Nonetheless, the number of the most \ha-massive galaxies is several orders of magnitude smaller than that of the less \ha-massive galaxies, therefore their impact on the logarithmic average of \ha masses is limited. 

To demonstrate that the break measured in this paper is not an artefact of the HI selection, we also build mock MIGHTEE-\ha samples with no break in the underlying $M_{\rm HI}-M_{\star}$ relation for LTGs, which is described by a single power law (i.e. our Model A). The mock samples are shown in the left panel of Figure~\ref{fig:mock_C}. The solid blue line is the best fitting of Model A for LTGs with the blue shaded areas being the 1$\sigma$ statistical uncertainties. By comparing the black circles with the black dashed lines, we find a similar result as in the previous section that the “observed” broken relation of the Model B mock LTGs with ETGs in the right panel agrees with the data better than that of the Model A mock LTGs  with ETGs in the left panel especially at around the transition mass range. The relative evidence between Models A and B for fitting the underlying $M_{\rm HI}-M_{\star}$ relation is $\ln(\mathcal{Z_B})$ - $\ln(\mathcal{Z_A}) = 4.4 \pm 0.4$, which is slightly larger than the Bayes factor of $3.9\pm 0.3$ when the LTGs and ETGs are considered as a whole. In other words, our Bayesian analysis suggests that modelling the LTGs with a break $M_{\rm HI}-M_{\star}$ relation is favoured by our MIGHTEE-\ha observations. We note that our analysis shows that the actual position of the break and the slope of the relation below and above the break are challenging to constrain, as can be inferred from the error bars in Table~\ref{tab:post-hi-C}. The full MIGHTEE survey, and a combination of MIGHTEE with other \ha surveys, such as Looking At the Distant Universe with the MeerKAT Array \citep[LADUMA; ][]{Blyth2016} will provide much stronger constraints.

We also note that there are other selection effects, such as the limited volume meaning we are susceptible to different environments, which may impact on our measurement of the underlying $M_{\rm HI}-M_{\star}$ relation. However, these effects are likely to be subdominant with respect to the flux-limited nature in our sample, and hard to be quantified without the help of large numerical simulations including dwarf \ha galaxies with their masses down to $\log_{10}(M_{\rm HI}$/$M_{\odot}) \sim 7$. We also cannot create a mock spiral galaxy population with our approach to assess their intrinsic $M_{\rm HI}-M_{\star}$ relation due to the uncertain correlation between their morphology, stellar and gas components. We look forward to seeing these aspects in the future hydro-dynamical and semi-analytic galaxy simulations.

\section{Conclusions}
\label{sec:conclusions}

We have developed a Bayesian technique that allows us to measure the $M_{\rm HI}-M_{\star}$ relation above or below the detection threshold in a unified way while taking into account its intrinsic scatter without binning the datasets. We implement this technique with the MIGHTEE-\ha Early Science data, and highlight our main results as:

\begin{itemize}
\item We model the upper envelope of the $M_{\rm HI}-M_{\star}$ relation down to $M_{\rm HI}\sim 10^7 M_{\odot}$, and up to z=0.084 using a \ha-selected sample of 249 galaxies. We use a double power law model to fit our data, and find this non-linear model is preferred by the data over the linear model, with a transition stellar mass of $\log_{10}(M_\star/M_{\odot}) = 9.15\pm0.87$, which roughly corresponds to the break in the stellar mass of $M_\star\sim10^{9} M_\odot$ found by \cite{Maddox2015}. Beyond this transition (or break) stellar mass, the slope of $M_{\rm HI}-M_{\star}$ relation flattens.

\item We also examine the corresponding SFR-$M_{\rm HI}$ relation and find that it is almost linear across the whole \ha mass range, albeit with a large scatter of $\sim0.48$ dex. Combined with the flattening feature on the $M_{\rm HI}-M_{\star}$ relation, this supports the hypothesis that the shortage of \ha gas supply is likely ultimately responsible for the quenching of the star formation activity observed in massive main sequence galaxies.

\item By separating our full sample into spirals, irregulars, mergers and ellipticals, we find the \ha sample is dominated by the spirals at the high mass end, and by the irregulars at the low mass end. These two type of galaxies exhibit significantly different slopes for the $M_{\rm HI}-M_{\star}$ relation, and are likely to be responsible for the detected transition stellar mass from the full sample, although we cannot rule out a pure mass dependence. In addition, we find that the ellipticals show a lower fraction of \ha mass than other types from the \ha-selected sample, and the highest mass galaxies show a higher fraction of \ha mass than predicted by hydrodynamic simulations \citep{Dav__2019}, although small number statistics prohibits a strong statement about the \ha characteristics of elliptical galaxies and the most massive ones.

\item We created mock galaxies above and below the MIGHTEE-\ha detection threshold with two broad galaxy populations of late- galaxies and early-type galaxies to measure the underlying $M_{\rm HI}-M_{\star}$ relation over the last billion years. We find that the \ha selection can lift the "observed" $M_{\rm HI}-M_{\star}$ relation on average but the turnover feature on this relation is largely immune to this effect, albeit with a weaker break, regardless of whether the two galaxy populations are taken as a whole or separately in their intrinsic bivariate distribution of \ha and stellar masses.

\item We fit a linear underlying $M_{\rm HI}-M_{\star}$ scaling relation (i.e. Model A) to the observed relation from our MIGHTEE-\ha observation in addition to the non-linear underlying relation (i.e. Model B). Although both models can mimic a broken $M_{\rm HI}-M_{\star}$ relation for our MIGHTEE-\ha observation, the Bayesian evidence suggests that the non-linear model is strongly favoured by our data over the linear one. This fact indicates that a careful analysis is needed to establish whether the observed knee in the $M_{\rm HI}-M_{\star}$ scaling relation  is real or not.

\item We also find that the evidence for a break in the intrinsic underlying $M_{\rm HI}-M_{\star}$ relation of LTGs is stronger than the evidence for a break in the upper envelope of spirals, demonstrating that the underlying break is stronger than the observed break for the same/similar galaxy samples. The evidence for a break is also stronger for LTGs and ETGs when modelled separately than as a whole sample in the underlying relation.

\end{itemize}

Taken together, our new analysis using the MIGHTEE-H{\sc i} Early Science data agrees with the results presented in \cite{Maddox2015}, where they also found an upper envelope in the amount of H{\sc i} that a galaxy can retain is dependent on its stellar mass, and we find that this is likely to be related to the morphology of the galaxy. A direct cause of this result could be the tight link between specific angular momentum (or halo spin parameter) and the gas fraction \citep{Obreschkow_2016,Kurapati_2021,Mancera_Pi_a_2021,Mancera2021, Hardwick_2022} for rotation-dominated galaxies. Interestingly, the transition mass that we find using our double-power law (Model B) to describe the upper envelope in the $M_{\rm HI} - M_{\star}$ relation corresponds to the $M_{\rm HI}/M_{\star}$ ratio at which we find that the spin axis of the galaxy to flip from aligned to mis-aligned from its nearest filament, using a subset of the same data \citep{Tudorache2022}. Given that \cite{Maddox2015} suggest that at $M_{\star} > 10^8$\,M$_\odot$, galaxies with higher \ha fractions sit in haloes with higher spin parameters which can work to stabilise \ha disks, the spin parameter may in turn be related to their proximity to a filament, along which the gas flows in towards the galaxy \citep[e.g.][]{Codis2018}. Given the limited statistics available in \cite{Tudorache2022} and this study, we cannot decisively investigate these multi-dimensional trends, however, with the full MIGHTEE survey such an analysis would be within reach.

\section*{Data availability}
The MIGHTEE-\ha spectral cubes and source catalogue will be released as part of the first data release of the MIGHTEE survey.

\section*{Acknowledgements}
The MeerKAT telescope is operated by the South African Radio Astronomy Observatory, which is a facility of the National Research Foundation, an agency of the Department of Science and Innovation. We acknowledge use of the IDIA data intensive research cloud for data processing. The IDIA is a South African university partnership involving the University of Cape Town, the University of Pretoria and the University of the Western Cape. The authors acknowledge the Centre for High Performance Computing (CHPC), South Africa, for providing computational resources to this research project.

We acknowledge the use of the ilifu cloud computing facility - \url{www.ilifu.ac.za}, a partnership between the University of Cape Town, the University of the Western Cape, the University of Stellenbosch, Sol Plaatje University, the Cape Peninsula University of Technology and the South African Radio Astronomy Observatory. The ilifu facility is supported by contributions from IDIA and the Computational Biology division at UCT and the Data Intensive Research Initiative of South Africa (DIRISA).

We thank the anonymous referees for their constructive comments that have improved this paper greatly. HP, MJJ, and MGS acknowledge support from the South African Radio Astronomy Observatory (SARAO) towards this research (\url{www.sarao.ac.za}). 
MJJ and AAP acknowledge generous support from the Hintze
Family Charitable Foundation through the Oxford Hintze Centre
for Astrophysical Surveys and the UK Science and Technology Facilities Council [ST/S000488/1]. IP acknowledges financial support from the Italian Ministry of Foreign Affairs and International Cooperation (MAECI Grant Number ZA18GR02) and the South African Department of Science and Technology's National Research Foundation (DST-NRF Grant Number 113121) as part of the ISARP RADIOSKY2020 Joint Research Scheme. SK is supported by the south african research chairs initiative of the department of science and technology and national research foundation. MB acknowledges support from the Flemish Fund for Scientific Research (FWO-Vlaanderen, grant G0G0420N). SHAR is supported by the South African Research Chairs Initiative of the Department of Science and Technology and National Research Foundation. RB acknowledges support from an STFC Ernest Rutherford Fellowship [grant number ST/T003596/1].

For the purpose of Open Access, the
author has applied a CC BY public copyright licence to any Author Accepted Manuscript version arising from this submission. 




\bibliographystyle{mnras}
\bibliography{references} 


\appendix



\bsp	
\label{lastpage}
\end{document}